\documentclass[preprint]{aastex}
\usepackage{natbib}
\usepackage{lineno}
\newcommand\pflux{\mbox{${\rm \, ph \,\, cm^{-2} \, s^{-1}}$}}
\shorttitle{Fermi LAT detected blazars}
\shortauthors{The Fermi LAT Collaboration}

\title{The Third Catalog of Active Galactic Nuclei Detected by the  {\em Fermi} Large Area Telescope}
\author{
M.~Ackermann\altaffilmark{1}, 
M.~Ajello\altaffilmark{2}, 
W.~B.~Atwood\altaffilmark{3}, 
L.~Baldini\altaffilmark{4}, 
J.~Ballet\altaffilmark{5}, 
G.~Barbiellini\altaffilmark{6,7}, 
D.~Bastieri\altaffilmark{8,9}, 
J.~Becerra~Gonzalez\altaffilmark{10,11}, 
R.~Bellazzini\altaffilmark{12}, 
E.~Bissaldi\altaffilmark{13}
% ...
}
\altaffiltext{1}{Deutsches Elektronen Synchrotron DESY, D-15738 Zeuthen, Germany}
\altaffiltext{2}{Department of Physics and Astronomy, Clemson University, Kinard Lab of Physics, Clemson, SC 29634-0978, USA}
\altaffiltext{3}{Santa Cruz Institute for Particle Physics, Department of Physics and Department of Astronomy and Astrophysics, University of California at Santa Cruz, Santa Cruz, CA 95064, USA}
\altaffiltext{4}{Universit\`a di Pisa and Istituto Nazionale di Fisica Nucleare, Sezione di Pisa I-56127 Pisa, Italy}
\altaffiltext{5}{Laboratoire AIM, CEA-IRFU/CNRS/Universit\'e Paris Diderot, Service d'Astrophysique, CEA Saclay, 91191 Gif sur Yvette, France}
\altaffiltext{6}{Istituto Nazionale di Fisica Nucleare, Sezione di Trieste, I-34127 Trieste, Italy}
\altaffiltext{7}{Dipartimento di Fisica, Universit\`a di Trieste, I-34127 Trieste, Italy}
\altaffiltext{8}{Istituto Nazionale di Fisica Nucleare, Sezione di Padova, I-35131 Padova, Italy}
\altaffiltext{9}{Dipartimento di Fisica e Astronomia ``G. Galilei'', Universit\`a di Padova, I-35131 Padova, Italy}
\altaffiltext{10}{NASA Goddard Space Flight Center, Greenbelt, MD 20771, USA}
\altaffiltext{11}{Department of Physics and Department of Astronomy, University of Maryland, College Park, MD 20742, USA}
\altaffiltext{12}{Istituto Nazionale di Fisica Nucleare, Sezione di Pisa, I-56127 Pisa, Italy}
\altaffiltext{13}{Istituto Nazionale di Fisica Nucleare, Sezione di Bari, 70126 Bari, Italy}
% ...

\begin{abstract}
The third catalog of active galactic nuclei (AGNs) detected by the {\it Fermi}-LAT  (3LAC) is presented. It is based on the third  {\it Fermi}-LAT catalog (3FGL) of sources detected between 100 MeV and 300 GeV with a Test Statistic ($TS$) greater than 25, between 2008 August 4 and 2012 July 31. The 3LAC includes 1591 AGNs  located at high Galactic latitudes ($|b|>10\arcdeg$), a  71\% increase over the second catalog based on 2 years of data. There are 28 duplicate associations, thus 1563 of the 2192 high-latitude gamma-ray sources of the 3FGL catalog  are AGNs.  Most of them (98\%) are blazars. About half of the newly detected blazars are of unknown type, i.e., they lack spectroscopic information of sufficient quality to determine the strength of their emission lines. Based on their gamma-ray spectral properties,  these sources are evenly split between flat-spectrum radio quasars (FSRQs) and BL~Lacs. The most abundant detected BL~Lacs are of the high-synchrotron-peaked (HSP) type. About 50\% of the BL~Lacs have no measured redshifts. A few new rare outliers (HSP-FSRQs and  high-luminosity HSP BL~Lacs)  are reported.  The general properties of the 3LAC sample confirm previous findings from earlier catalogs. The fraction of 3LAC blazars in the total population of blazars listed in BZCAT remains non-negligible even at the faint ends of the BZCAT-blazar radio, optical and X-ray flux distributions, which is a clue that even the faintest known blazars could eventually shine in gamma rays at LAT-detection levels. The energy-flux distributions of the different blazar populations are in good agreement with extrapolation from earlier catalogs.
\end{abstract}

\keywords{gamma rays: observations --- galaxies: active --- galaxies: jets --- BL Lacertae objects: general}

\begin{document}
\enlargethispage*{1000pt}

\section{Introduction}

Since its launch in 2008, the {\it Fermi}-LAT has revolutionized our knowledge of the gamma-ray sky above 100 MeV.
\pagebreak
 Its unique combination of high sensitivity, wide field of view, large energy range, and nominal sky-survey operating mode  has enabled a complete mapping and continuous monitoring of the gamma-ray sky to an unprecedented level.  
Several catalogs or source lists, both general and specialized (AGNs, pulsars, supernova remnants, pulsar wind nebulae, gamma-ray bursts, very-high-energy candidates) have already been produced. These constitute important resources to the astronomical community. The successive AGN lists and catalogs, LBAS \citep[LAT Bright AGN Sample,][]{LBAS}, 1LAC \citep[][]{1LAC} and  2LAC \citep[][]{2LAC,2LACErr}, first and second LAT AGN catalogs respectively, have triggered numerous population studies \citep[e.g.,][]{Ghi09,Pad12,Mey12,Ghi12,Aje12,Fin13,Ghi13,Gio13,Dab12,Mas12}, provided suitable samples, e.g., to probe the Extragalactic Background Light \citep[EBL, ][]{Abdo_EBL,
  EBL_Science}, offered suitable target lists to investigate the dichotomy between  gamma-ray loud and gamma-ray quiet blazars at other wavelengths \citep{Lis09,Kov09,Ojh10,Lis11,Pin12,paperPLANCK}, and served as references for works on individual sources \citep[e.g.,][]{Abr13,Tav13}.

This paper presents the  third catalog of AGNs detected by the {\it Fermi}-LAT  after four years of operation (3LAC).  It is a follow-up of the 2LAC \citep{2LAC} and makes use of the results of the 3FGL catalog \citep{3FGL}, a sequel to the 2FGL catalog \citep{2FGL}. The latter contained 1873 sources. In addition to dealing with more data, the 3FGL benefits from improved data selection, instrument response functions and analysis techniques.  The 3FGL catalog includes 3033 sources with  a Test Statistic\footnote{We use the Test Statistic $TS = 2\Delta\log L$ for quantifying how significantly a source emerges from the background, comparing
the likelihood function $L$ with and without that source.}  ($TS$) greater than 25.   Among them,  2192 sources are detected at $|b|>10\arcdeg$ where $b$ is the Galactic latitude. Among these 2192, 1563 (71\%) are associated with high confidence with 1591 AGNs, which constitute the 3LAC.  The 3LAC represents a sizeable improvement over the 2LAC as it includes 71\% more sources (1591 vs. 929\footnote{See \cite{2LACErr} for a 2LAC Erratum. The corrected 2LAC full and clean samples include 929 and 827 sources, respectively.   A total of 63 of the 88 sources mistakenly included in the initial 2LAC full sample  are now in the 3LAC catalog.}) with an updated data analysis.       

The paper is organized as follows.  In Section 2, the observations by the LAT and the analysis employed to produce
the four-year catalog are described.  In Section 3, we explain the methods for associating
\mbox{gamma-ray} sources with AGN counterparts and the different schemes for
classifying 3LAC AGNs. Section 4  provides a brief census of the 3LAC sample and discusses sources of particular interest.
Section 5 summarizes some of the properties of the 3LAC, including the
\mbox{gamma-ray} flux distribution, the \mbox{gamma-ray} spectral properties, the
redshift distribution, the \mbox{gamma-ray} luminosity distribution and  the \mbox{gamma-ray} variability properties.  In
Section 6, we address the connection with populations of blazars detected in the two neighboring energy bands, namely the hard X-ray and very-high-energy (VHE) bands.  We discuss the implications of the 3LAC results in Section 7
and present our conclusions in Section 8.

In the following, we use a $\Lambda$CDM cosmology with values from the {\it Planck} results
\citep{Planckcosmo}; in particular, we use $h = 0.67$, $\Omega_m = 0.32$, and
$\Omega_\Lambda = 0.68$, where the Hubble constant $H_0=100h$~km~s$^{-1}$~Mpc$^{-1}$.

\section{\label{sec:obs}Observations with the Large Area Telescope --- Analysis Procedures}

The gamma-ray results used in this paper were derived in the context of the 3FGL catalog, so we only briefly summarize the  analysis here and we refer the reader to the paper describing the 3FGL catalog \citep{3FGL} for details. No additional analysis of the gamma-ray data was performed in the context of the present paper except for the fitting of the monthly light curves described in Section \ref{sec:var}. The broadband SED fitting described in Section  \ref{sec:sedclass} was also carried out in this work. 

The data were collected over the first 48 months of the mission, from 2008 August 4 to 2012 July 31 (MJD 54682 to 56139). Time intervals during which the rocking angle of the LAT was greater than 52$^{\circ}$ were excluded and a cut on the zenith angle of gamma rays of 100$^{\circ}$ was applied to limit the contribution of Earth-limb gamma rays. Time intervals with bright gamma-ray bursts and solar flares were excised. The reprocessed Pass7REP\_V15\_Source event class  was used, with photon energies between 100 MeV and 300 GeV.  This event class shows a narrower point-spread function above 3 GeV than the Pass7\_V6\_Source class used in 2FGL. 
The source detection procedure started with an initial set of sources from the 2FGL analysis; not just those reported in that catalog, but also including all candidates failing the significance threshold. With these seeds, an all-sky likelihood analysis produced an ``optimized'' model, where parameters characterizing the diffuse components\footnote{The Galactic diffuse model and isotropic background model (including the gamma-ray diffuse and residual charged-particle backgrounds) are described in the 3FGL paper. Alternative Galactic diffuse models were tested as well.}, in addition to sources were fitted. The analysis of the residual $TS$ map provided new seeds that were included in the model for a new all-sky likelihood analysis. This iterative procedure yielded over 4000 seeds that were then passed on to the maximum likelihood analysis for source characterization. 

Events from the front and back sections of the LAT tracker \citep[see ][ for details]{atwood} were treated separately in the analysis. The analysis was performed with the binned likelihood method below 3 GeV and the unbinned method above 3 GeV. These methods are implemented in the  {\it pyLikelihood} library  of the Science Tools\footnote{http://fermi.gsfc.nasa.gov/ssc/data/analysis/documentation/Cicerone/} (v9r23p0).   Different spectral fits were carried out with a single power-law function (\mbox{$dN/dE=N_0\:(E/E_{0})^{-\Gamma}$}) and  a log-parabola function \citep[\mbox{$dN/dE=N_0\:(E/E_{0})^{-\alpha-\beta \log(E/E_0)}$},][]{Mas04}, where $N_0$ is a normalization factor, $\Gamma$, $\alpha$ and $\beta$ are spectral parameters and $E_0$ is an arbitrary reference energy adjusted on a source-by-source basis to minimize the correlation between $N_0$ and the other fitted parameters over the whole energy range (0.1 to 300 GeV). Whenever the difference in log(likelihood) between these two fits was greater than 8 (i.e., $TS_{curve}$, which is defined as twice this difference,  was greater than 16), the log-parabola results were retained. 
For 3C 454.3, an exponentially cutoff power law (\mbox{$dN/dE=N_0\:(E/E_{0})^{-\Gamma}\:\exp [(E_{0}/E_{c})^b-(E/E_{c})^b]$},
 where  $E_c$ is the cutoff energy and $b$ the exponential index) was needed to provide a reasonable fit to the data.
The photon spectral index ($\Gamma$) was obtained from the single power-law fit for all sources. A threshold of $TS=25$, as calculated with the power-law model, was applied to all sources, corresponding to a significance of approximately 4 $\sigma$. At the end of this procedure, 3033 sources survived the $TS$ cut and constitute the 3FGL catalog. 

Power-law fits were also performed in five different energy bands (100 to 300 MeV; 300 MeV to 1 GeV; 1 GeV to 3 GeV; 3 GeV to 10 GeV; 10 GeV to 300 GeV), from which the energy flux was derived.  A variability index (TS$_{VAR}$) was constructed from a likelihood test based on the monthly averaged light curves,  with the null (alternative) hypothesis corresponding to the source being steady (variable). A source is identified as being variable at the 99\%  confidence level if the variability index is equal or greater than 72.44, $TS_{VAR}$ being distributed as a $\chi^2$ function with 47 degrees of freedom.   

Some of the 3FGL sources were flagged as doubtful when certain issues arose during their analysis (see 3FGL for a full list of these flags).  The issues that most strongly affected the 3LAC list are: i) sources with  $TS >$ 35 going down to $TS <$ 25 when changing the diffuse model, ii)  photon flux ($>$ 1 GeV) or energy flux ($>$ 100 MeV) changed by more than 3 $\sigma$ and 35\% when changing the diffuse model, iii) sources located close to a brighter neighbor  (the conditions are defined in Table 3 of 3FGL), and iv) source Spectral\_Fit\_Quality $>$ 16.3 ( Spectral\_Fit\_Quality is the $\chi^{2}$ between the fluxes in five energy bands and the spectral model).  We developed a clean selection of sources by excluding sources that have any of the 3FGL analysis flags set. About 91\% (1444/1591) of the 3LAC sources survived this cut. Although the Spectral\_Fit\_Quality condition may reject sources with unusual spectra, this condition ensures that the spectral properties discussed in the following are not affected by analysis issues.

\begin{figure}
\centering
\resizebox{13cm}{!}{\rotatebox[]{0}{\includegraphics{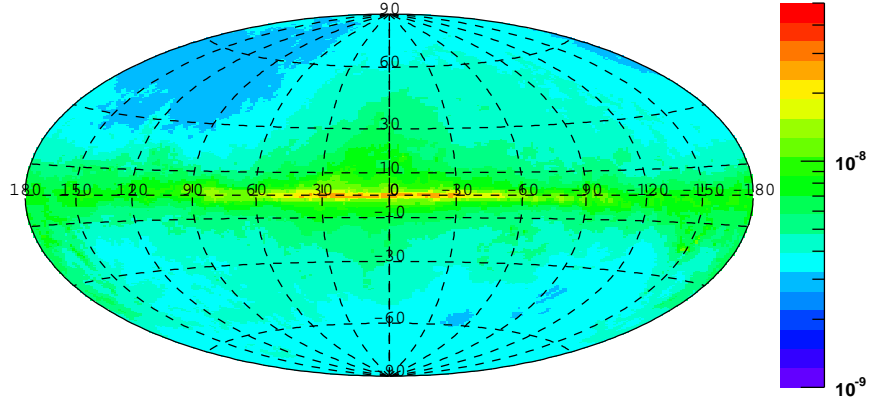}}}
\caption{Point-source flux limit in units of \pflux{} for $E > 100$~MeV and photon spectral index $\Gamma = 2.2$ as a function of sky location (in Galactic coordinates) for the 3LAC time interval.}
\label{fig:sensitivity}
\end{figure}

\begin{figure}
\centering
\plottwo{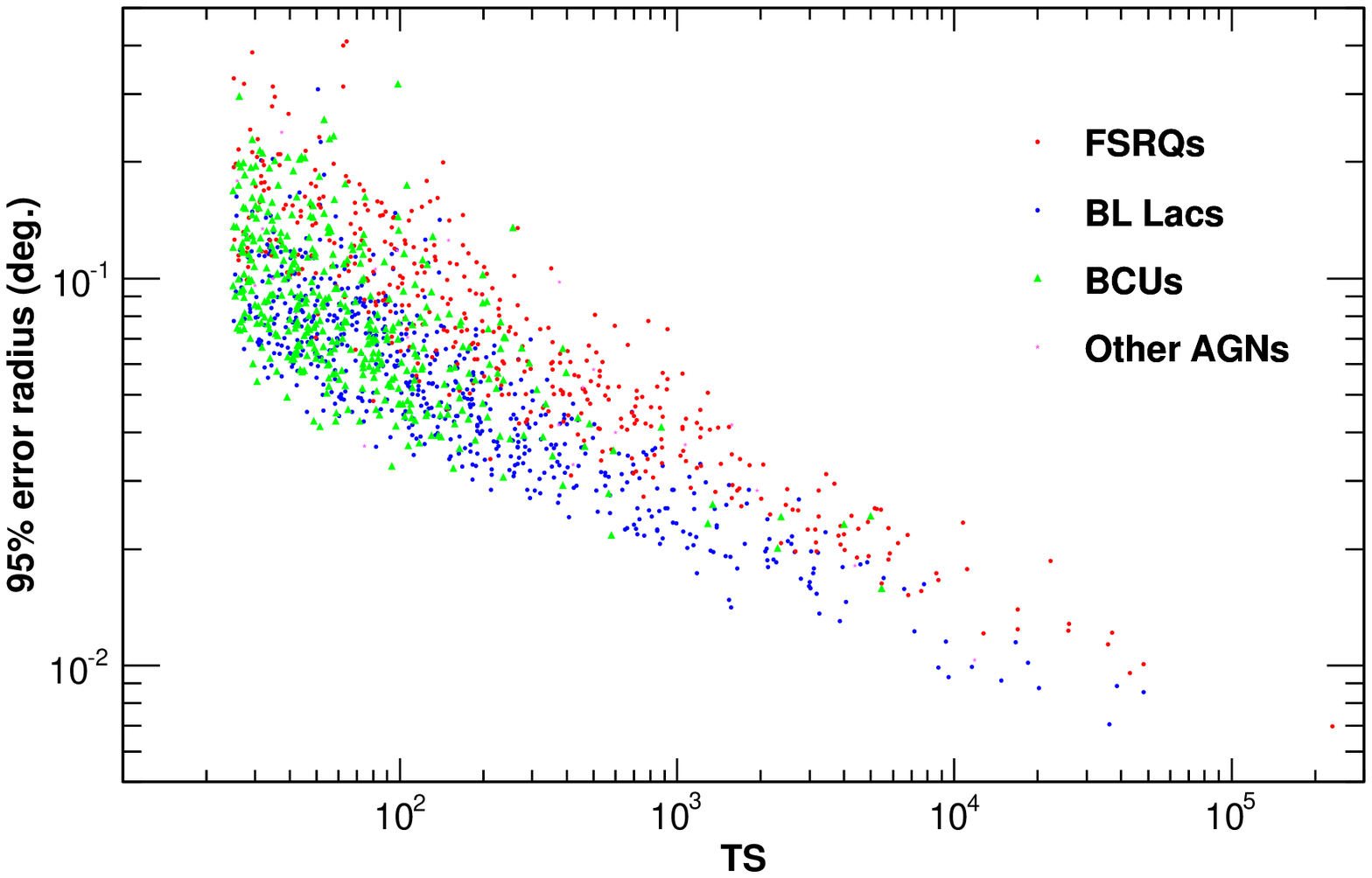}{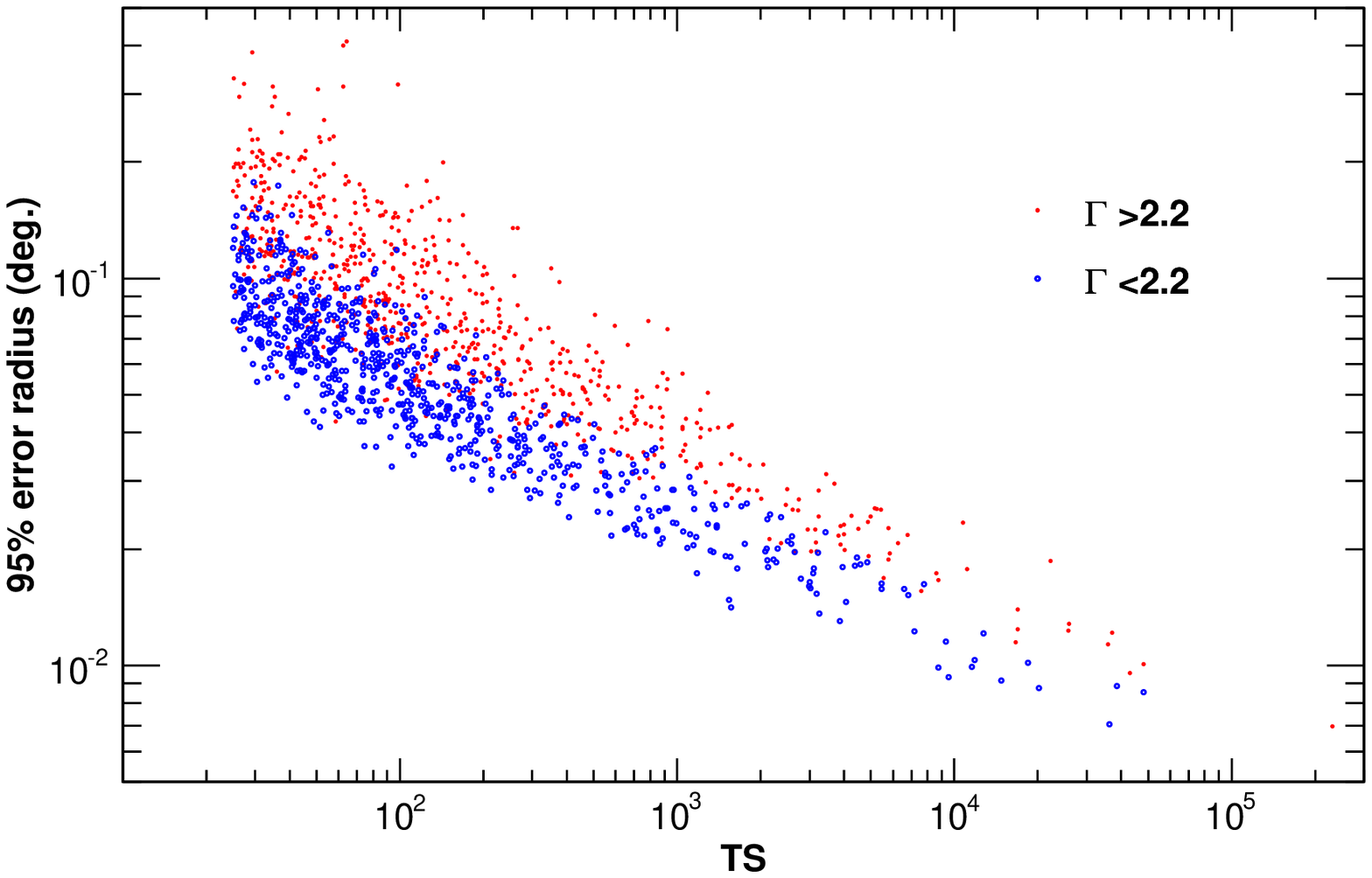}
\caption{ 95\% error radius versus  $TS$. Left: Red circles: FSRQs, blue circles: BL~Lacs, green triangles: unknown type (BCUs). Right: Sources with $\Gamma>2.2$ (red) and $\Gamma<2.2$ (blue).}
\label{fig:r95_TS}
\end{figure}

 A  map of the LAT flux limit, calculated for the four-year period covered by this catalog, a  $TS=25$ and a photon index of 2.2, is shown in Galactic coordinates in Figure \ref{fig:sensitivity}. A map computed for a photon index of 1.8 would look very similar, with flux limits about four times lower.  
The 95\% error radius, $\theta_{95}$, defined as the geometric mean  of the semi-major and semi-minor axes of the source location ellipse (see 3FGL),  is plotted as a function of $TS$ in Figure \ref{fig:r95_TS}. It ranges from about $0\fdg007$ for 3C~454.3, the brightest LAT blazar, to $0\fdg08$-$0\fdg3$ for sources just above the detection threshold depending on the gamma-ray spectral slope.    

\section{Source Association and Classification}

In this work we look for candidate counterparts to 3FGL gamma-ray sources based on positional association with known cataloged objects that display AGN-type spectral characteristics. These characteristics are a flat radio spectrum between 1.4 GHz and 5 GHz, an AGN-like broadband emission, core compactness or radio extended emission.

We recall here that in the context of AGNs, {\it identification}  is only firmly established when correlated variability with a counterpart detected at other energies has been reported. So far, only 26 AGNs have met this condition (see 3FGL). For the rest, we use statistical approaches to find associations between LAT sources and AGNs. We will refer to the so-associated AGNs as the counterparts, although identification is not strictly established.

We apply the Bayesian Association Method \citep{1FGL} to catalogs of sources that were already classified and/or characterized. These catalogs  come from specific instruments providing information on the spectrum and/or broadband emission. If a catalog reports an AGN classification, that is used.  Otherwise the classification is made according to the criteria described below.

%This powerful method, however, cannot deal with large all-sky surveys, which would be too computationally intensive.  Therefore, t
To broaden the possibility of associating a candidate AGN knowing  its broadband emission characteristics, we added the Likelihood Ratio Method \citep{2LAC}.  This method can handle large uniform all-sky surveys and take the source space-density distribution into account. In the case of general radio or X-ray surveys,  including AGN and non-AGN sources,  the classification procedure is the same as for the Bayesian Association Method.

These two association approaches have been extensively described in previous catalog papers, so only updates will be given here (see Sect. \ref{sec:assoc}).   

\subsection{\label{sec:classif}Source Classification}

To define the criteria that a source must fulfill to be considered as an AGN, the ingredients are primarily the optical spectrum and to a lesser extent other characteristics such as radio loudness, flat/steep radio spectrum between 1.4 GHz and 5 GHz, broadband emission, flux variability, and polarization.

We stress that we are classifying the candidate counterpart to a 3FGL source. If available,  the earlier classification in the literature of each reported candidate counterpart was checked.

\subsubsection{\label{sec:optclass}Optical Classification}

To optically classify a source we made use of different resources, in decreasing order of precedence: optical spectra from our intensive follow-up program \citep{Sha13}, the BZCAT list \citep[ i.e., classification from this list, which is a compilation of sources ever classified as blazars, ][]{bzcat}, and spectra available in the literature,  e.g. SDSS \citep{sdss}, 6dF \citep{6df}, when more recent than the version 4.1.1 of BZCAT (August 2012). The latter information was used only if we found a published spectrum. % The relevant references are reported in the electronic table.

The resulting classes are as follows:

\begin{itemize}
\item confirmed classifications:
\subitem flat-spectrum radio quasar (FSRQ), BL~Lac, radio galaxy, steep-spectrum radio quasar (SSRQ), Seyfert,  and Narrow-Line Seyfert 1 (NLSy1) -- these are sources with a well-established classification in the literature and/or through a well evaluated optical spectrum (with clear evidence for or lack of emission lines).
\item tentative classifications:
\subitem BCU -- {\it blazar candidates of uncertain type}: these are considered candidate blazars because the association methods (see  Sections \ref{sec:gtsrcid} and  \ref{sec:like}) select a candidate counterpart that satisfies at least one of the following conditions:
\subsubitem $a$) a BZU object (blazar of uncertain/transitional type) in the BZCAT list;
\subsubitem $b$) a source with multiwavelength data in one or more of the {\it WISE} \citep{WISE},
 AT20G \citep{AT20G_CAT}, VCS \citep{vcs}, CRATES \citep{crates}, PMN-CA \citep{pmn-ca}, CRATES-Gaps \citep{crates}, or CLASS \citep{class} source lists, that indicates a flat radio spectrum, and shows a typical two-humped, blazar-like spectral energy distribution (SED);
\subsubitem $c$) a source included in radio and X-ray catalogs not listed above and for which we found a typical two-humped, blazar-like SED \citep[see ][]{Bot07}.

The {\it BCU} sources are divided into three sub-types:
\subsubitem {\it BCU I}: the counterpart has a published optical spectrum but not sensitive enough for a classification as an FSRQ or a BL Lac;
\subsubitem {\it BCU II:} the counterpart is lacking an optical spectrum but a reliable evaluation of the SED synchrotron-peak position is possible;
\subsubitem {\it BCU III}: the counterpart is lacking both an optical spectrum and an estimated  synchrotron-peak position but shows blazar-like broadband emission and a flat radio spectrum;

\subitem AGN -- the counterparts show SEDs typical of radio-loud compact-core objects, but data are lacking in the literature to be more specific about their classes.

\end{itemize}

\subsubsection{\label{sec:sedclass}SED Classification}

To better characterize the candidate counterparts of the 3FGL sources which we consider candidate blazars or more generally radio-loud AGNs, we studied their broadband spectral energy distributions by collecting all data available in the literature\footnote{We made extensive use of the SED Builder on-line tool available at the ASI Science Data Center, http://tools.asdc.asi.it/SED/.}.

We use the estimated value of the (rest-frame) broadband-SED  synchrotron peak frequency $\nu_\mathrm{peak}^\mathrm{S}$ to classify the source as either a low-synchrotron-peaked blazar (LSP, for sources with $\nu_\mathrm{peak}^\mathrm{S} < 10^{14}$~Hz), an intermediate-synchrotron-peaked blazar (ISP, for $10^{14}$~Hz~$< \nu_\mathrm{peak}^\mathrm{S} < 10^{15}$~Hz), or a high-synchrotron-peaked blazar (HSP, if $\nu_\mathrm{peak}^\mathrm{S} > 10^{15}$~Hz). We refer the reader to the 2LAC paper for the list of broadband data used in this procedure.

The estimation of $\nu^S_\mathrm{peak}$ relies on  a 3rd-degree polynomial fit of the low-energy hump of the SED performed on a source-by-source basis, while in previous catalogs (1LAC, 2LAC) an empirical parametrization of the SED based on the broadband indices $\alpha_{ro}$  (radio-optical) and $\alpha_{ox}$  (optical-X-rays) was used
\citep[see][]{SEDpaper}.  In this new method, some sources changed SED classification with respect to the 2LAC (see below). 

This new procedure allows more objects to be assigned peak parameters than the empirical method since there is no need of a measured X-ray flux if the curvature is sufficiently pronounced  in the IR-optical band. Even though a scrupulous check was performed for each individual source, caution is advised in using these $\nu^S_\mathrm{peak}$ values determined using non-simultaneous broadband data.  Significant contamination from thermal/disk radiation may result in  overestimation of the $\nu^S_\mathrm{peak}$  values of  FSRQs, while  the contribution of the host galaxy may bias the peak estimate towards lower frequencies in BL~Lacs.  Comparing the two procedures  indicates that the new procedure leads to an average shift of +0.26 (rms: 0.49) and $-$0.05 (rms: 0.64) in $\log \nu_\mathrm{peak}^\mathrm{S}$ relative to the previous one for FSRQs and BL~Lacs respectively, which we take as typical systematic uncertainties.  

In the electronic tables, we report the so-obtained observer-frame values of  $\nu^S_\mathrm{peak}$, as well as the rest-frame  values (i.e., corrected by a $(1+z)$ factor). For BL~Lac and BCU sources without measured redshifts, a redshift $z=0$ was assumed for the SED classification, but we omit these sources in figures showing $\nu^S_\mathrm{peak}$. Assuming a redshift of 1 for these sources as suggested by \cite{Gio13} would lead to a shift in the rest frame $\log \nu_\mathrm{peak}^\mathrm{S}$ of +0.3, taken as an additional systematic uncertainty.

\begin{figure}
\centering
\resizebox{10cm}{!}{\rotatebox[]{0}{\includegraphics{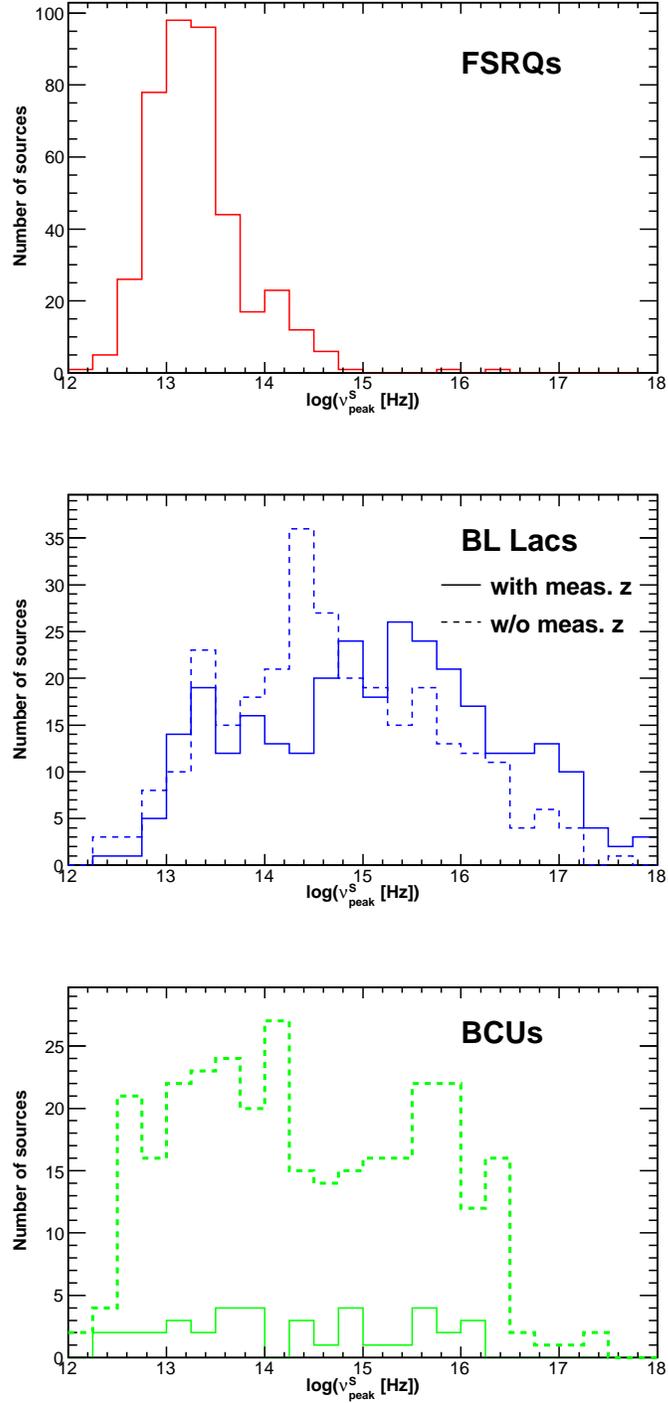}}}
\caption{Distributions of the synchrotron peak frequency $\nu^S_{peak}$ for FSRQs (top), BL~Lacs (middle) and BCUs (bottom) in the Clean Sample (defined in Section \ref{sec:assres}). The solid and dashed histograms correspond to sources with and without measured redshifts, respectively. The (1+z) correction factor (to convert to rest-frame values)} has thus  been applied to $\nu^S_{peak}$ only for the former.
\label{fig:syn_hist}
\end{figure}

The $\nu_\mathrm{peak}^\mathrm{S}$  distributions for FSRQs and BL Lacs are displayed in Figure \ref{fig:syn_hist}. The FSRQ distribution is sharply peaked around  $\log \nu_\mathrm{peak}^\mathrm{S}$=13 while BL~Lacs span the whole parameter space from low (LSP) to the highest frequencies (HSP). The BCU distribution  resembles that of BL~Lacs with an additional fairly weak component akin to FSRQs at this low $\nu_\mathrm{peak}$ end.

\subsection{\label{sec:assoc}Source Association}
\subsubsection{\label{sec:gtsrcid}The Bayesian Association Method}

This method \citep[see ][]{1FGL} uses Bayes' theorem to calculate the posterior probability that a catalog source is the true  counterpart of a LAT source.  The significance of a spatial coincidence between a candidate counterpart from a catalog $C$ and a
LAT-detected \mbox{gamma-ray} source is evaluated by examining the local density of counterparts from $C$ in the vicinity of the LAT source. 
If the candidate counterpart has not been established as an AGN in a catalog C, all we have is a positional association. The nature of the candidate counterpart is subsequently studied through the literature and SED study (See Sect. \ref{sec:classif}). The catalogs used in 3LAC are the 13th edition of the Veron catalog \citep{AGNcatalog}, version 4.1.1 of BZCAT \citep{bzcat}, the CRATES and CGRaBs catalogs \citep{crates}, the 2010 December 5 version of the VLBA Calibrator Source List\footnote{
The VLBA Calibrator Source List can be downloaded from http://www.vlba.nrao.edu/astro/calib/vlbaCalib.txt.}, 
the most recent version of the TeVCat catalog\footnote{http://tevcat.uchicago.edu}, and the Australia Telescope 20-GHz Survey  \citep[AT20G; ][]{AT20G_CAT}, which contains entries for 5890 sources observed at declination $\delta <$0$\arcdeg$. Associations with the {\it Planck} Early Release Catalogs \citep{Planckcatalog} were performed as well, but an association solely with a {\it Planck} counterpart was not considered sufficient to call the source an AGN candidate as {\it Planck} detects sources of various types. Additions relative to 2LAC are the list of {\it WISE} gamma-ray blazar candidates from \cite{WISE} and  \cite{Ars14}. The whole list of catalogs used in this method is given in Table 12 of the 3FGL paper \citep{3FGL}.  

%,Mas14

\subsubsection{\label{sec:like}The Likelihood-Ratio Association Method}

The Likelihood Ratio method (LR) has frequently been used to assess identification probabilities for radio, infrared and optical sources \citep[e.g., ][]{deRuiter77,Pre83,ss92,Lon98,Mas01,2LAC}. It is based on uniform surveys in the radio and in X-ray  bands, enabling us to search for possible counterparts among the faint radio and X-ray sources. The LR makes use of counterpart densities (assumed spatially constant over the survey region) through the $\log N - \log S$ relation and therefore the source flux.
As for the Bayesian method applied to catalogs without classification information, we can only claim a positional association  for these counterparts. The nature of the candidate counterpart is subsequently studied through the literature and SED properties (see Sect. \ref{sec:classif}).

We made use of a number of relatively uniform radio surveys.  Almost all radio AGN candidates of possible interest are in the  NRAO VLA Sky Survey  \citep[NVSS; ][]{NVSScatalog}, and the Sydney University Molonglo Sky Survey \citep[SUMSS; ][]{SUMSScatalog}. We also added AT20G.  In this way we are able to look for radio counterparts with detections at higher frequencies. To look for additional possible counterparts we cross-correlated the LAT sources with the most sensitive all-sky X-ray survey, the {\sl ROSAT} All Sky Survey (RASS) Bright and Faint Source Catalogs \citep{RASSbright,RASS_FAINT_CAT}. 
 The method, which computes the probability that a suggested association is the ``true'' counterpart, is described in detail in Section  3.2 of the 2LAC paper.
A source is considered as a likely counterpart of the gamma-ray source if its reliability, $\log LR$,  (see Eq. 4  in the 2LAC paper) is greater than 0.8 in at least one survey. The critical values of $\log LR_c$ above which the reliability is  greater than $0.8$ are  1.69, 0.52, 2.42 and 5.80 for the NVSS, SUMSS, RASS, and AT20G surveys respectively.

\subsection{\label{sec:assres} Association Results}

The adopted threshold for the association probability is 0.80 in either method.
This value represents a compromise between association efficiency and purity. 
  As in previous LAC catalog versions, we define a Clean Sample as 3LAC single-association sources free of the analysis issues mentioned in Section \ref{sec:obs}.
Table \ref{tab:assoc} compares the performance of the two methods in terms of total number of associations, estimated number of false associations $N_{false}$, calculated  as $N_{false}=\sum\limits_i (1-P_i)$, where $P_i$ is the association probability for the $i$th source,  and number of sources  associated solely via a given method, $N_S$, for the full and Clean samples. 

The fraction of sources associated by both methods is 71\% (1150/1591), 379 and 62 sources being solely associated with the Bayesian and LR methods respectively.  Among the former, 177 sources are associated due to the list of {\it WISE} gamma-ray blazar  candidates only (over 1000 3FGL sources have counterparts in that catalog).  
 The overall false-positive rate is 1.9\%. The estimated number of false positives among the 571 sources not previously detected in 2FGL and previous LAT catalogs is 12.0 (2.1\%).

\begin{figure}
\centering
\resizebox{14cm}{!}{\rotatebox[]{0}{\includegraphics{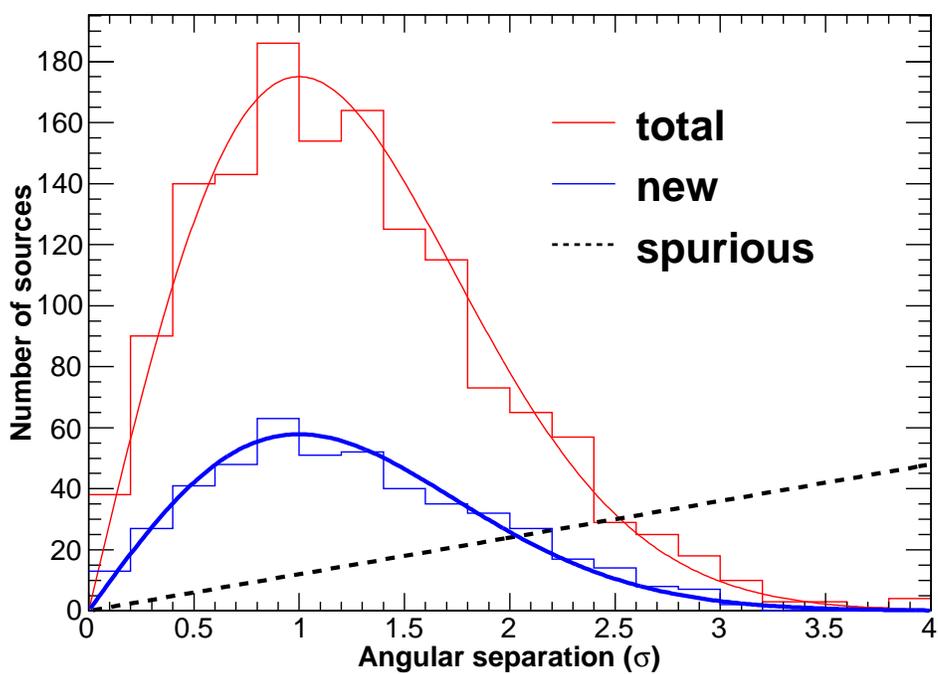}}}
\caption{Distributions of  normalized angular separation between 3LAC sources and their assigned counterparts. The normalization factor $\sigma$ is defined in the text. Red: total. blue: new sources. The curves correspond to the expected distribution for real associations, the dashed line illustrates that expected for spurious associations.}
\label{fig:separation}
\end{figure}

Figure \ref{fig:separation} displays the distributions of separation distance between the gamma-ray sources and their assigned counterparts, normalized to $\sigma=\theta_{95}/\sqrt{-2\log(0.05)}$, for the whole sample and for the newly detected sources. Both agree well with the  distributions expected for real associations, as expected from the overall low false-positive rate.

\subsection{Blazar candidates by the Australia Telescope Compact Array}

In this section, we point out blazar candidates derived from the recent work of \cite{petrov13} but not all included in 3LAC.  Using the Australia Telescope Compact Array (ATCA) at 5 GHz and 9 GHz, \cite{petrov13} detected 424 sources  in the LAT error ellipses of southern unassociated 2FGL sources. They found that  84 of them have radio-source counterparts with a spectral index flatter (i.e., greater) than $-$0.5.  

The  424 sources are characterized by weak radio fluxes ($< 100$ mJy), and were thus missing from the previous AT20G. Flat spectrum radio sources cannot be directly associated with  extragalactic sources like blazars, as peculiar Galactic objects (like, for example, $\eta$ Carinae, microquasars, compact HII regions, planetary nebulae) can also exhibit a flat radio spectrum. On the other hand a steep radio spectrum does not rule out an extragalactic nature. A total of 24 sources among the 84 flat-spectrum ones are included in 3LAC as they now fulfill the required criterion (association probability greater than 0.8). An additional 21 sources listed in Table \ref{tab:petrov} show double-humped radio-to-gamma-ray SEDs resembling those of blazars of unknown type (BCU) but they have association probabilities below threshold.  More data may help secure these associations in the future.

\section{\label{sec:cat}The Third LAT AGN Catalog (3LAC)}

\subsection{\label{sec:census} Census}

Table \ref{tab-census} summarizes the 3LAC source statistics. The 3LAC includes 1591 objects with 467 FSRQs, 632 BL~Lacs, 460 BCUs and 32 non-blazar AGNs. Their properties are given in  Table \ref{tab:clean}. 

 A total of 1563 gamma-ray sources have been associated with radio-loud AGNs among 2192 $|b|>10\arcdeg$ 3FGL sources, corresponding to an overall association fraction of 72\%. The fraction changes substantially between the northern and southern celestial hemispheres (843/1136=74\% and 731/1056=69\% respectively), an effect essentially entirely driven by unassociated southern-hemisphere BL~Lacs as discussed below.  

Only sources in the Clean Sample will be used in the following in tallies and figures unless stated otherwise. It includes 1444 objects with 414 FSRQs, 604 BL~Lacs, 402 BCUs and 24 non-blazar AGNs.  

A comparison of the results inferred from the 3LAC and 2LAC enables the following observations:
\begin{itemize}
\item The 3LAC Clean Sample includes 619  more sources than the 2LAC Clean Sample, i.e., a  75\% increase.  Of these, 477 sources are new (81 FSRQs, 146 BL~Lacs, 240 blazars of  unknown type, 10 non-blazar objects); the other sources were present in previous {\it Fermi} catalogs but not included in Clean Samples for various reasons (e.g., the corresponding  gamma-ray sources were not associated with AGNs, had more than one counterpart or were flagged in the analysis).  The fraction  of new sources (not present in 1FGL or 2FGL) is slightly higher for hard-spectrum (i.e., $\Gamma<$2.2) sources than for soft-spectrum ones (i.e., $\Gamma>$2.2), 51\% vs. 47\% respectively, but the relative increase reaches 72\% for very hard-spectrum (i.e., $\Gamma<$1.8) sources.  
\item The fraction of blazars of unknown type (BCU) has increased notably between the two catalogs (from 20\% to 28\%). The number of these sources in the 3LAC Clean Sample has increased by more than a factor of 2.5  relative to that in the 2LAC Clean Sample, being almost equal to the number of FSRQs. This increase is mainly due to the lower probability of having a published high-quality spectrum available for these fainter sources because of the lack of  optical/near-infrared observing programs. The census of the BCU sources in the Clean Sample is: 49 BCU I, 308 BCU II, 45 BCU III.
\item The relative increase in BCUs drives a drop in the proportions of FSRQs and BL~Lacs, which only represent 29\% and 41\% of the 3LAC Clean Sample respectively  (38\% and 48\% for 2LAC). The relative increase in the number of sources with respect to 2LAC is 34\% and 42\% for FSRQs and BL~Lacs respectively. 
\item Out of 827 sources in the 2LAC Clean Sample, a total of 69 are missing in the 3LAC Clean Sample (42 in the full sample), some of them probably due to variability effects.  A few others are present in 3FGL but with shifted positions, ruling out their association with their former counterparts.
\end{itemize}

\begin{figure}
\centering
\epsscale{2}
\plottwo{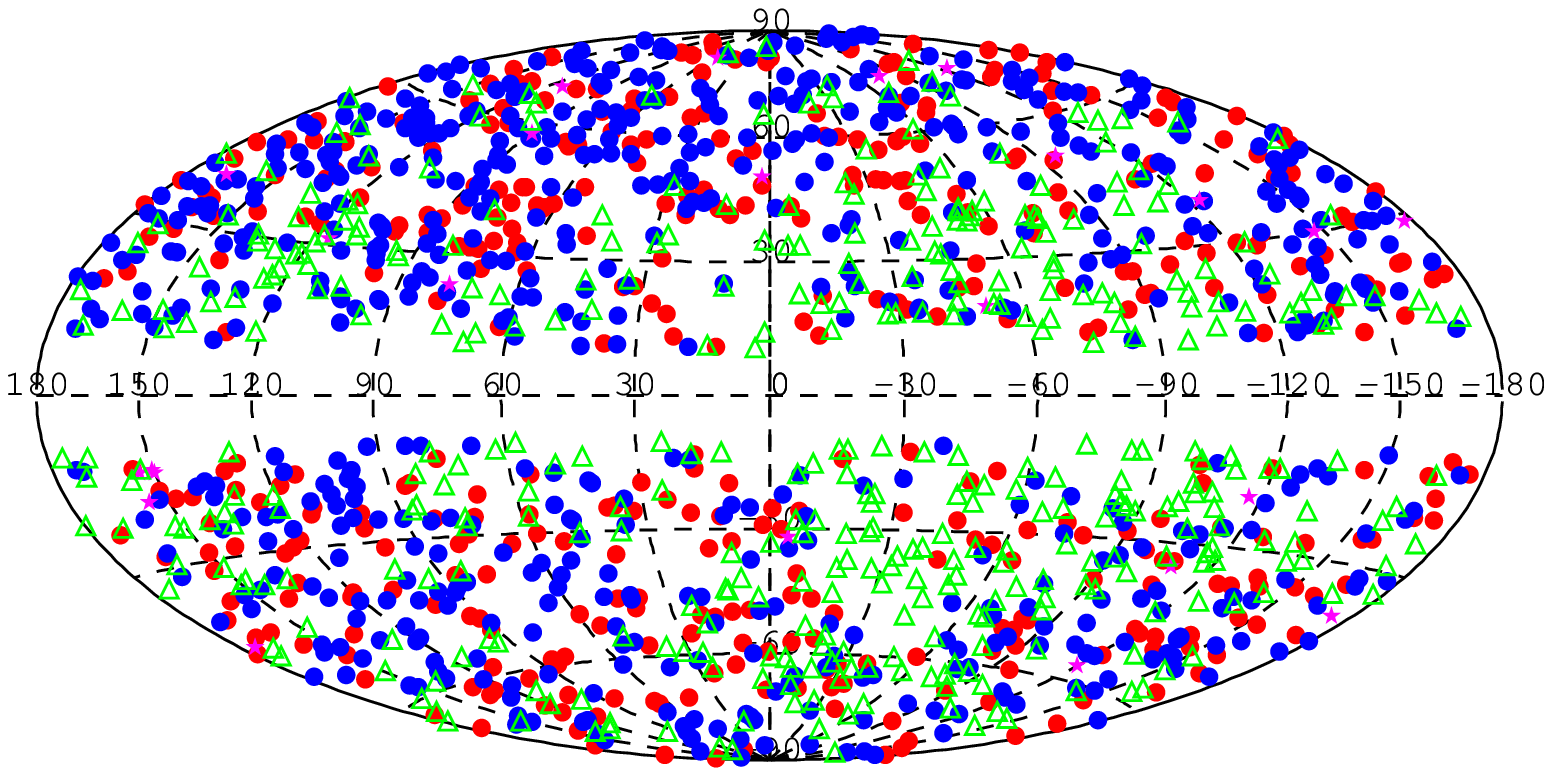}{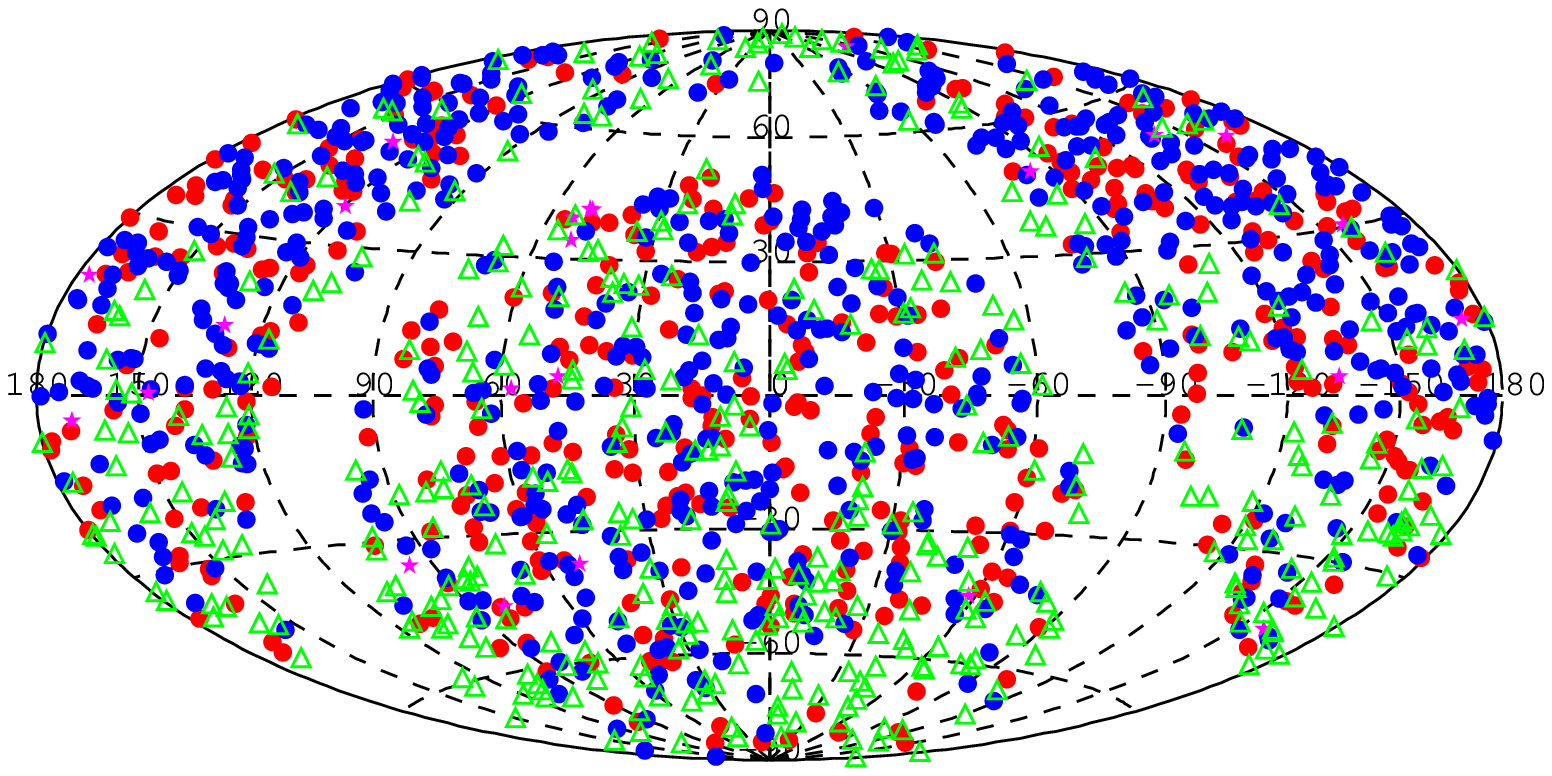}
\caption{Locations of the sources in the Clean Sample in Galactic (top) and  J2000 equatorial (bottom) coordinates.  Red circles:\ FSRQs, blue circles:\ BL~Lacs, green triangles: blazars of unknown type,  magenta stars: other AGNs.}
\label{fig:sky_map}
\end{figure}

The loci of sources in the Clean Sample are shown in Figure \ref{fig:sky_map}, both in Galactic and celestial coordinates.  The deficit in classified AGNs in the region of the celestial south pole  already reported in 2LAC is clearly visible, while a relative  excess is seen in the region of the celestial north pole. This anisotropy is mainly driven by BL~Lacs, with 51\% more sources in the northern Galactic hemisphere (362) than in the southern one (242). This effect is ascribed to the  relative incompleteness of the counterpart catalogs in the southern hemisphere (for instance NVSS only covers the $\delta>$-40$\arcdeg$ sky, where  $\delta$ is the declination). It is very partially offset by an observed relative excess (54 sources) of associations with blazars of unknown type in the south relative to the north.

\subsection{\label{sec:misaligned}Non-Blazar Objects and Misaligned AGNs}

Blazars represent the overwhelming majority of 3LAC AGNs, with non-blazar AGNs only constituting 2\% of the sample. In 2LAC, eleven sources were classified as AGNs, i.e., were neither confirmed blazars nor blazar candidates (such as BCUs).  Although there may have been evidence for their flatness in radio emission or broadband emission, our intensive optical follow-up program did not provide clear evidence for optical blazar characteristics. Nine of them remain in 3LAC, and are now all classified as BCUs, except for one now classified as a BL Lac.

Misaligned AGNs (MAGNs), with jets pointing away from the observer, are not favored GeV sources. By MAGNs we mean radio-loud AGNs with jets directed at large angle relative to the line-of-sight that display steep radio spectra ($\alpha_{r} \ge$ 0.5, with the usual convention that $F_{\nu} \propto \nu^{-\alpha}$ ) and bipolar or quasi-symmetrical structures in radio maps. The larger jet inclination angle relative to blazars means the observed radio emission from the relativistic jet is not significantly Doppler boosted, making it less prevalent over other radio components  such as synchrotron radiation from mildly relativistic outflows or extended radio lobe emission \citep{magn}.

Table~\ref{tab:magn} summarizes the non-blazar objects and MAGNs in the 3FGL/3LAC, noting also their previous appearances in the 2FGL/2LAC and 1FGL/1LAC. All the 1FGL sources, detected in 11 months of exposure, were subsequently studied with 15 months of data \citep{magn}.

M~87 was one of the first new {\it Fermi}-LAT detections \citep{M87} of a source classified as a
non-blazar object, being a low-power \citet{fan74} type-I
(FRI) radio galaxy. Many of the newly-associated non-blazar objects are nearby FRIs -- J0758.7+3747 (3C~189, a.k.a., B2~0755+37) and 3C~264. The gamma-ray detection of the latter case was recently
reported in a study of its parent cluster Abell 1367 \citep{cluster}, although the gamma rays likely
originate from the AGN. 
We remark that 0.8$\arcmin$ away from 3C 189 lies the quasar SDSS J075825.87+374628.7 with redshift 1.50. With the resolution of the NVSS, this source cannot be disentangled from the radio emission of 3C 189. This may be the reason why this source is not present in the NVSS catalog, precluding the estimation of the association probability with the gamma-ray source.

NGC~1275 (3C~84, Perseus A) 
was first detected in the initial LAT bright source list based on 3 months of data \citep{bsl}. It was probably detected previously with COS-B \citep{str83}, but not with EGRET. In the {\it Fermi} era, it is a strong source exhibiting GeV variability \citep{ngc1275,kat10}. 
3C~120 is not listed in any of the FGL catalogs but its detection was reported in a 15-month study \citep{magn}. There are indications that 3C~120 undergoes a series of flares with a low long-term average flux. For instance, in September 2014 a flaring source positionally consistent with 3C~120 was detected with a high significance  \citep[$TS>$50; ][]{ATel6529}. The closest 3FGL source, 3FGL  J0432.5+0539, lies $0\fdg35$ away with an 95\% error radius of ~$0\fdg15$, hampering association with 3C~120 by our methods. This gamma-ray source has a soft spectrum ($\Gamma=2.7\pm 0.1$) comparable with that ascribed to 3C~120 \citep{magn,kat11}. The possibility of two separate, soft-spectrum sources cannot be excluded.  Another known example from previous lists is 3C~78 \citep[NGC 1218;][]{1LAC}.

\begin{figure}
\centering
\resizebox{9cm}{!}{\rotatebox[]{0}{\includegraphics{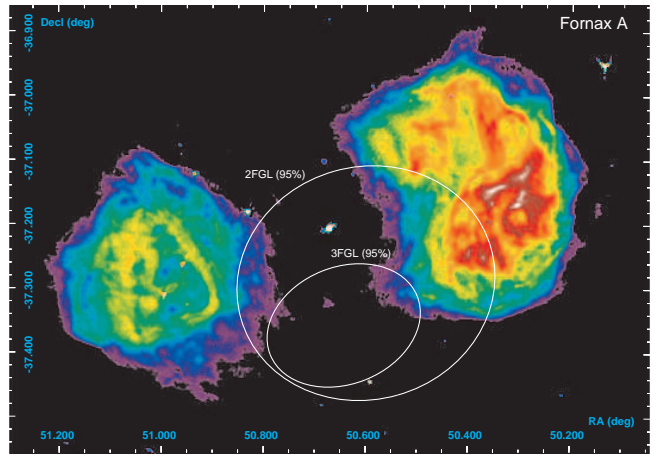}}}
\caption{VLA image of Fornax A at 1.5 GHz \citep{Fom89}. The 95\% error ellipses of the closest 3FGL and 2FGL sources are overlaid. 
%The 2FGL ellipse was more overlaid on the radio emission, but the 3FGL one has drifted away. The 2FGL and 3FGL sources are associated.
}
\label{fig:ForA}
\end{figure}

Cen~A was also reported in the initial LAT bright source list \citep{bsl}, confirming the
EGRET source \citep{3EGcatalog,sre99}. It remains the only AGN with a significant detection of extended gamma-ray emission
\citep{cena}.  There is no convincing case of extended emission in other radio galaxies with relatively large radio extensions, such as Cen~B
\citep{kat13}, NGC~6251 \citep{tak12}, and Fornax A. Fornax A may be a good case to investigate this emission \citep{che07,geo08}. The closest 3FGL source is offset from the Fornax A core by 0$\fdg$15, while the 95\%-contour distance is 0$\fdg$092 (see Figure \ref{fig:ForA} for a VLA 1.5 GHz image).  NGC~6251 (one square degree in solid angle) was also detected by EGRET \citep{muk02}. Its location shifted between 1LAC and 2LAC towards the western radio lobe.

3C~111 was also a previous EGRET source \citep{har08} and also shows apparent variability \citep[e.g.,][]{magn,kat11,gra12}. It joins the other two FR type-II sources listed in Table~\ref{tab:magn}:  3FGL J1442.6+5156 (3C~303) and
3FGL J0519.2$-$4542 (Pictor A).  The latter are also broad-lined radio galaxies (BLRGs), and are new detections. The LAT detection of Pictor A was reported by \cite{bro12}\footnote{The results of \cite{bro12} are in tension  with those presented here, with their reported statistical uncertainties being very small given the low source flux.} following a previous tentative detection \citep{kat11}.  

The previous LAT detections of PKS~0625$-$35 and IC~310, two radio galaxies with BL~Lac  characteristics, were  reported in 2LAC, and are confirmed. IC~310 has been classified as a head-tail galaxy \citep{ner10}, but recent works have found increasing evidence for blazar-like properties, e.g.,  blazar-like VLBI jet structure \citep{kad12} and extremely fast TeV variability \citep{Ale14}. The source 4C +39.12 (3FGL J0334.2+3915) was classified as a low-power compact source by \cite {Gio01}, 
separate from its Fanaroff-Riley classification. Two new compact steep-spectrum (CSS) sources are detected:  3FGL J1330.5+3023 (3C~286) and 
3FGL J0824.9+3916 (4C +39.23B). While both are CSS, the latter is a duplicate  association (the other association being the FSRQ blazar  
4C+39.23) so is not in the Clean Sample. The former has the morphology of a medium symmetric object (MSO), like that of the LAT-detected FSRQ 4C +55.17 \citep{mcc11}.

The gamma-ray detections of 3C~207 and 3C~380 were first reported in 1LAC.  They appear in the 3CRR catalog \citep{Lai83} by virtue of their bright low-frequency emission due to the presence of kpc-scale extended steep-spectrum radio lobes, thus are formally classified as steep-spectrum radio quasars (SSRQs). However, they contain pronounced flat-spectrum radio cores with superluminal motions measured in their parsec-scale jets, indicating that they are the most well-aligned sources to our line of sight amongst the SSRQs in the 3CRR  \citep[e.g., ][]{Wil91,Hou13,Lis13}. New ones to highlight are 3C~275.1 (3FGL J1244.1+1615), TXS 0348+013 (3FGL J0351.1+0128) and  4C +39.26 (3FGL J0934.1+3933). The SSRQ 4C +04.40 is part of a double association (with the FSRQ MG1 J120448+0408) of 3FGL J1205.4+0412.  

GB~1310+487 is a gamma-ray/radio-loud narrow-line AGN at z = 0.638, showing  a gamma-ray flare in November 2009  and located behind the disk of an unrelated emission-line galaxy at z = 0.500 \citep{Sok14}. 

Circinus, a type-2 Seyfert galaxy located at b=$-$3$\fdg$8 and thus not in 3LAC, was recently detected \citep{hay13}. Other Seyfert detections were investigated \citep{ten11,latseyferts,Len10}, but were found to be starburst galaxies \citep{latstarbursts}.

The detections of NGC~6951 (classified as a Seyfert 2 galaxy and a LINER, reported in 1LAC but missing in 2LAC), 3C~407 (a source with broad emission lines but with a fairly steep radio spectrum and reported in 2LAC), and NGC 6814 (type 1.5 Seyfert galaxy, also reported in 2LAC) are not confirmed. The same conclusion applies to  PKS~0943$-$76 \citep[studied in ][]{magn}. The previous claim that it has a FRII morphology was based on a low-resolution radio map from \citet{bur06}. The offset between the 4 year source and PKS 0943$-$76 is $0\fdg22$ while the radius of the source location region at the 95\% confidence level is $0\fdg12$. ESO 323$-$G77 (type 2 Seyfert galaxy), and  PKS0943$-$76 (radio galaxy), both reported in 2LAC, were actually both mis-associated because of an error in the LR association method \citep{2LACErr}. 

Five sources are associated with NLSy1. Four of them were  included in 2LAC:  3FGL J0325.2+3410 (BZU J0324+3410),  3FGL J0948.8+0021 (PMN J0948+0022), 3FGL J1505.1+0326 (BZQ J1505+0326), 3FGL J2007.8$-$4429 (BZQ J2007$-$4434), while 3FGL 0849.9+5108 (SBS 0846+513) was first reported by \cite{don11} and further studied by \cite{Dam12,Dam13}.

\subsection{\label{sec:indiv} Noteworthy Sources}

The highest redshift reported in 2LAC for an HSP-BL~Lac was 0.7. The 3LAC lists  seven (six in the Clean Sample)  HSP-BL~Lacs with redshifts greater than 1, six (five in the Clean Sample) of which were included in 2LAC but with other classifications or redshifts. They are briefly discussed below.
 
{\sl 3FGL~J0008.0$+$4713} is associated with MG4~J000800$+$4712. The redshift reported in 2LAC was 0.28 and its SED classification was LSP. \cite{Sha13} derived a redshift of 2.1 from the clear onset of the Lyman-$\alpha$ forest and their new procedure for estimating SED class together with {\it WISE} data classified this source as a HSP.       

{\sl 3FGL~J0630.9$-$2406} is associated with TXS~0628$-$240, an HSP-BL~Lac  for which $z \ga 1.238$ was determined from certain absorption features by \cite{Lan12}.

{\sl 3FGL~J1109.4+2411} is associated with 1ES~1106$+$244 and new spectroscopy from SDSS changed the redshift to 1.220.

{\sl 3FGL~J1312.5$-$2155} is associated with PKS~1309$-$216. In \cite{Sha13} a plausible Mg II feature is found; this single-line identification is in a small allowed redshift range ($z\simeq 1.6$). However previous data \citep{bzcat} show a questionable redshift of 1.491.

{\sl 3FGL~J2116.1+3339} is associated with B2~2114$+$33. The redshift quoted in 2LAC was 0.35, but a recent measurement by \cite{Sha13}  gives $z=1.596$ identifying a significant broad emission feature  with C IV, consistent with a weak bump in the far blue at Ly $\alpha$. A lower  redshift is possible if the purported  Ly $\alpha$ line is not real.

The newly detected source is {\sl 3FGL~J0814.5+2943},  associated with FBQS~J081421.2$+$294021 at $z=1.084$ \citep[from SDSS DR3, ][]{sdss}.

The highest redshift BL~Lac object is {\sl 3FGL~J1450.9$+$5200} associated with BZB~J1450$+$5201 with redshift $z=2.41$ coming from   new observations in \cite{Sha13}. The presence of the Ly-$\alpha$ forest  can suppress a part of the optical spectrum resulting in ISP classification, so probably the intrinsic synchrotron peak position is greater than our estimate.

 A low-redshift source, reported as a BCU in BZCAT, has recently been classified  as an FSRQ by  \cite{Gra14}: SBS 1646+499 (3FGL~J1647.4+4950) with $z=0.0467$.

Two HSP-FSRQs have been detected: BZB J0202+0849 (3FGL J0202.3+0851) and  NVSS J025037+171209 (3FGL J0250.6+1713) with LAT spectral photon indices of  2.05$\pm$0.16 and 1.98$\pm$0.19 respectively. 3FGL J0202.3+0851 was classified as a BL~Lac in 1LAC but new observations from \cite{Sha13} led to a reclassification  as an FSRQ.  These objects are probably transitional objects that show broad lines in the optical band when the continuum is low \citep[see, e.g., ][]{Rua14}.

\subsection{\label{sec:lowlat}Low Galactic Latitude AGNs}

Because of the intrinsic incompleteness of the counterpart  catalogs in this sky area ($|b|<10\arcdeg$), these sources are treated separately and are not included in the 3LAC or in the  analyses presented in the rest of the paper. 
We report associations for 182 blazars (75\% more than in 2LAC) located at $|b|<10\arcdeg$ : 24 FSRQs, 30 BL Lacs, 125 BCUs and 3 non-blazar AGNs. They are listed in Table \ref{tab:lowlat}. Extrapolating from the number of high-latitude sources and assuming the same sensitivity, about 340 sources would be expected in this area.   The discrepancy between expected and actual source numbers stems from the dual effect of a higher detection threshold due to a higher Galactic diffuse emission background (see Figure \ref{fig:sensitivity}) and a higher incompleteness of the counterpart  catalogs for this area.

\subsection{Comparison with 1LAC and 2LAC}

The revised 2LAC sample  \citep{2LACErr} includes 929 sources, 65 of which are missing in 3LAC (Table \ref{tab:missing}). Most do not make the $TS$ cut over the 4 year-long period, probably mainly due to variability. On the other hand, 56 unassociated sources in the 2FGL are now associated with blazars, thanks to a more complete set of counterpart catalogs and more precise localizations for the gamma-ray sources (arising from greater statistics and an improved instrument point-spread function). A total of 27 1FGL sources (not necessarily all in 1LAC) that were not listed in 2LAC are now included in 3LAC.   Some 51 2LAC sources have changed classifications in 3LAC, mostly due to improved data: 8 AGNs into BCUs, 1 AGN into a BL Lac, 39 BCUs into 34 BL~Lacs and 5 FSRQs, one  FSRQ into a BL~Lac (TXS~0404+075) and two BL~Lacs into FSRQs (B2~1040+24A and 4C~+15.54).

\subsection{Flaring Sources Detected in the Flare-Advocate Service}

The 3LAC catalog lists sources detected with high significance during 48 months.  Some blazars
flare during a limited time only and may be missing in 3LAC. If bright enough, some of them are
caught in near-real time by the {\it Fermi} Flare Advocate service, also known as Gamma-ray Sky Watcher (FA-GSW),
which we briefly describe here.

A day-by-day review of the whole gamma-ray sky, both by a human-in-the-loop and
by automated science processing (ASP) analysis \citep[see, e.g.,][]{Chi12}, results in the calculation of preliminary source
fluxes, tentative localizations and counterpart associations for any significant source detection.
This service is an important resource for the scientific community by providing alerts on
flaring or transient sources and by producing seeds for
follow-up variability and multiwavelength\footnote{For example the LAT MW Coordinating Group page: \texttt{confluence.slac.stanford.edu/x/YQw}}  studies 
\citep[see, e.g., ][]{ciprini2013}.

Since the beginning of the mission, daily reports are compiled internally to the Collaboration, while information and news are communicated via the LAT-MW mailing-list\footnote{Address: \texttt{lists.nasa.gov/mailman/listinfo/gammamw/}}, published in The Astronomer's Telegrams (ATels\footnote{ \texttt{www-glast.stanford.edu/cgi-bin/pub\_rapid}}), special GCN notices\footnote{ \texttt{gcn.gsfc.nasa.gov/gcn/fermi\_lat\_mon\_trans.html}}, and weekly summaries in the {\it Fermi} Sky Blog\footnote{The {\it Fermi} Gamma-ray Sky Blog:\\ \texttt{fermisky.blogspot.com}}. A total of 201 ATels were posted on behalf of the LAT Collaboration in the 48-month period considered in the 3FGL/3LAC, specifically from 2008 July 24 (the first ATel\#1628) to 2012 July 29 (ATel\#4285), primarily derived from the FA-GSW service. Some 143 ATels contained alerts and preliminary results about blazars and other AGN targets\footnote{An interactive incremental list is available at: $\texttt{www.asdc.asi.it/feratel/
 }$.} referring to 71 different FSRQs, 18 different BL Lac objects and 9 other AGNs or BCUs detected in flaring, hardening or enhanced activity states. Only one, PKS 1915$-$458 (z=2.47, ATel\#2666 and ATel\#2679) is not listed in the 3FGL/3LAC or in previous LAT catalogs. This high-redshift FSRQ  appears to only emit gamma rays  sporadically within short time intervals.

In addition three LAT sources announced in  ATels  and not present in the 3LAC  might have extragalactic source associations: {\it Fermi} J0052+1110  located at high Galactic latitude), PMN J1626$-$2426 (FSRQ in the vicinity of 3FGL J1626.2$-$2428 but outside its error ellipse and located behind an HII region), and PMN J0623$-$3350 (flat spectrum radio source reported as {\it Fermi} J0623$-$3351). A fourth LAT ATel source tentatively associated there with the FSRQ PKS 2136$-$642 is listed as 3FGL J2141.6$-$6412 in 3FGL but is now associated with the BCU PMN J2141$-$6411 that is separated $\sim 15'$ from the former.

\section{Properties of 3LAC Sources}

\subsection{\label{sec:flux}Flux and Photon Spectral Index}

\begin{figure}
\centering
\resizebox{14cm}{!}{\rotatebox[]{0}{\includegraphics{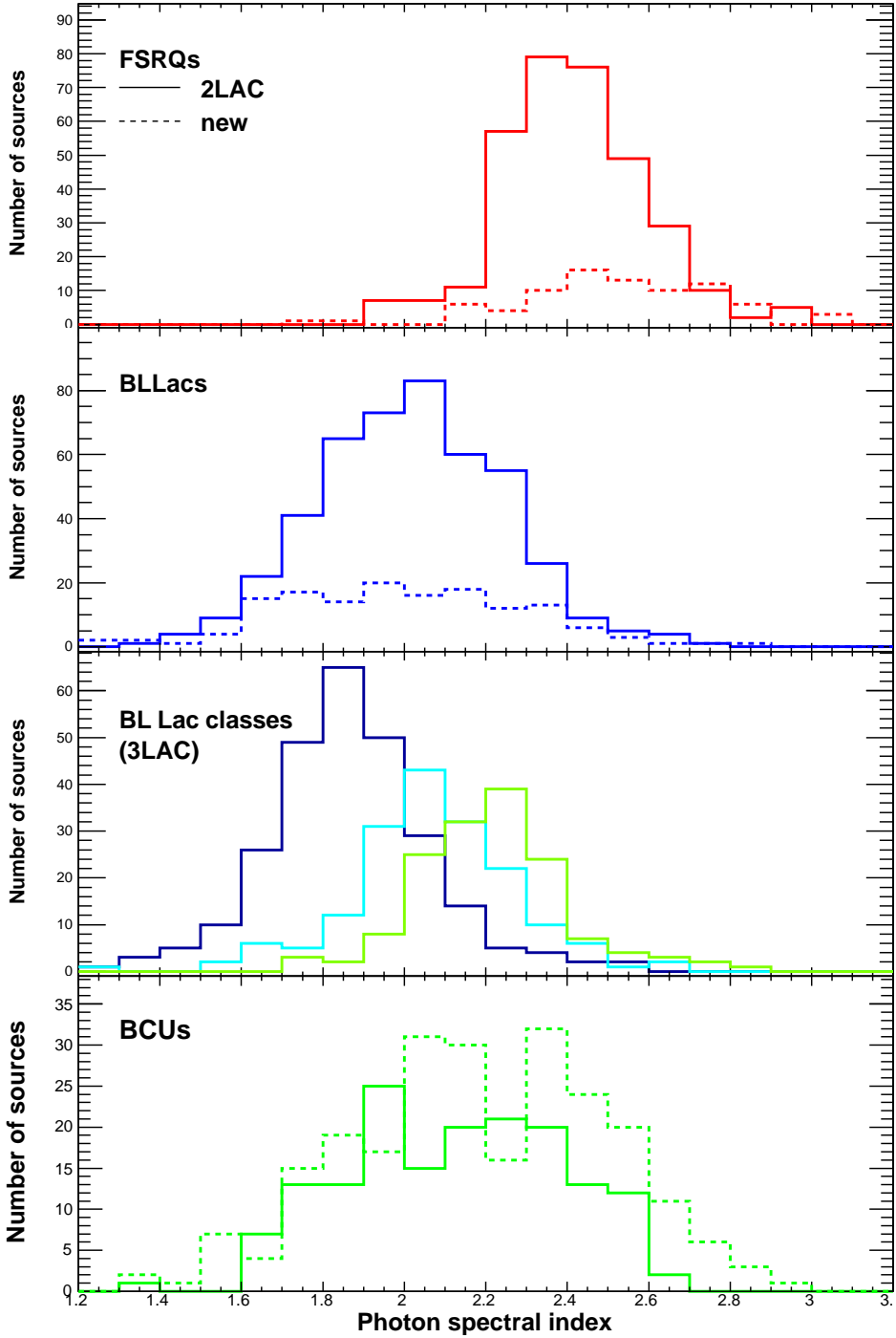}}}
\caption{Photon spectral index distributions. Top: FSRQs (solid: 2LAC sources, dashed: new 3LAC sources). Second from top: BL~Lacs (solid: 2LAC sources, dashed: new 3LAC sources). Third from top: 3LAC LSP-BL~Lacs (green), ISP-BL~Lacs (light blue), HSP-BL~Lacs (dark blue). Bottom:  blazars of unknown type (solid: 2LAC sources, dashed: new 3LAC sources).}
\label{fig:index}
\end{figure}

Figure \ref{fig:index} displays the photon index distributions for the different blazar classes both for the sources previously listed in 2LAC and the newly-detected sources. The newly-detected FSRQs are slightly softer than the 2LAC ones (2.53$\pm$0.03 vs. 2.41$\pm$0.01), indicating that the LAT  gradually detects more lower energy-peaked blazars. In contrast,  there is no significant spectral difference between the two sets of BL~Lacs. For BCUs, the distribution of the new sources extends further out on the high-index end ($\Gamma>$2.4), where the overlap with the BL~Lac distribution becomes very small.   The corresponding sources seem likely to be FSRQs.  

\begin{figure}
\centering
\resizebox{15cm}{!}{\rotatebox[]{0}{\includegraphics{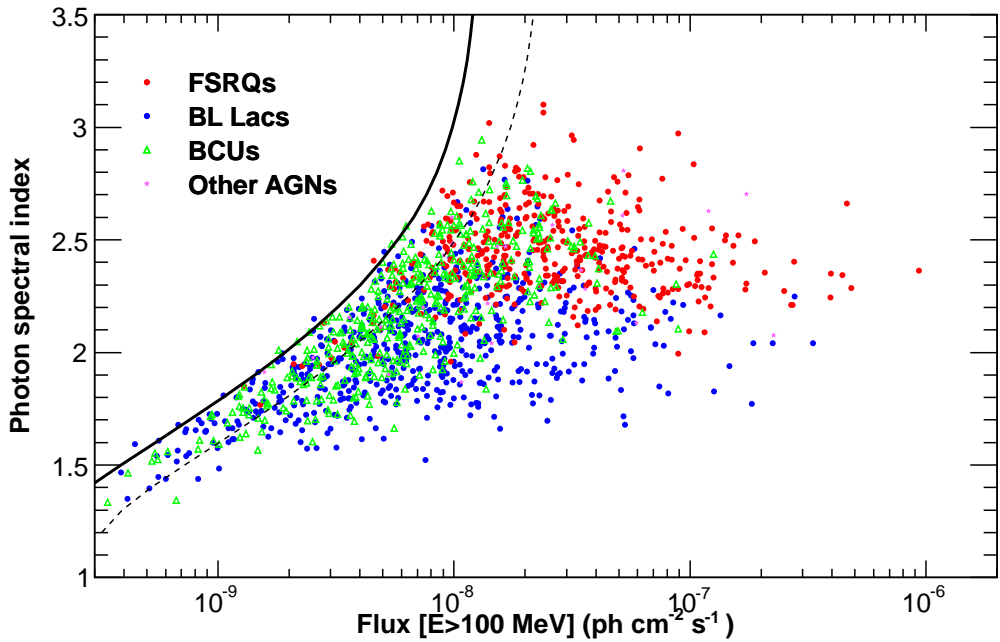}}}
\caption{Photon spectral index versus photon flux above 100 MeV for blazars in the Clean Sample. Red circles: FSRQs, blue circles:\ BL~Lacs, green triangles: blazars of unknown type, magenta stars:\ other AGNs. The { solid (dashed)} curve represents the approximate 3FGL (2FGL) detection limit based on a typical exposure.}
\label{fig:index_flux}
\end{figure}

\begin{figure}
\centering
\resizebox{15cm}{!}{\rotatebox[]{0}{\includegraphics{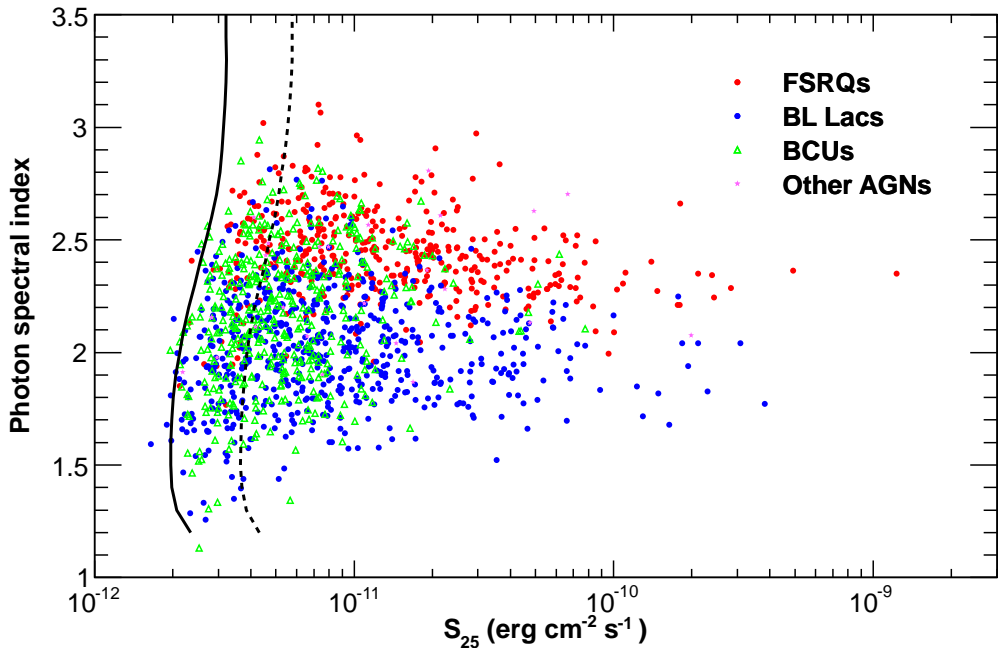}}}
\caption{Photon  spectral index versus energy flux between 100 MeV and 100 GeV, S$_{25}$. Red circles: FSRQs, blue  circles: BL~Lacs, green triangles: blazars of unknown type, magenta stars: other AGNs. The  solid (dashed) curve represents the approximate 3FGL (2FGL) detection limit  based on a typical exposure.}
\label{fig:index_S}
\end{figure}

Figures \ref{fig:index_flux} and \ref{fig:index_S}  show the photon index versus the photon flux and energy flux, respectively, together with estimated flux limits. As noted in 2LAC, the strong bias observed towards hard sources in the photon-flux limit essentially vanishes when considering the energy-flux limit above 100 MeV instead. (Note that this feature holds only for a lower bound of 100 MeV; other lower energy limits will bring about a dependence of the energy-flux limit on the spectral index).  

\begin{figure}
\centering
\resizebox{14cm}{!}{\rotatebox[]{0}{\includegraphics{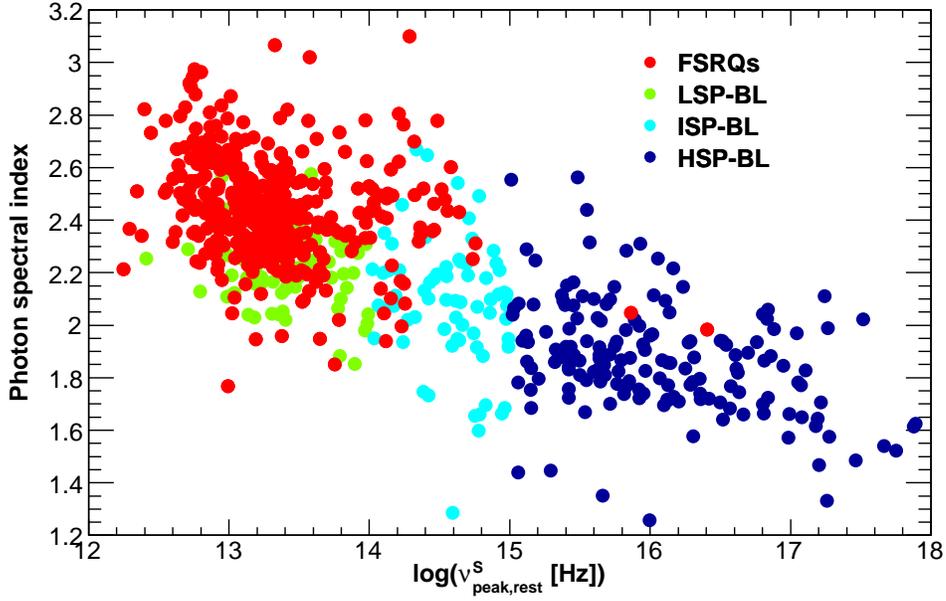}}}
\caption{Photon index versus frequency of the synchrotron peak $\nu^S_{peak,rest}$. Red: FSRQs, green: LSP-BL Lacs, light blue: ISP-BL Lacs,  dark blue: HSP-BL Lacs.}
\label{fig:index_nu_syn}
\end{figure}

\begin{figure}
\centering
\resizebox{14cm}{!}{\rotatebox[]{0}{\includegraphics{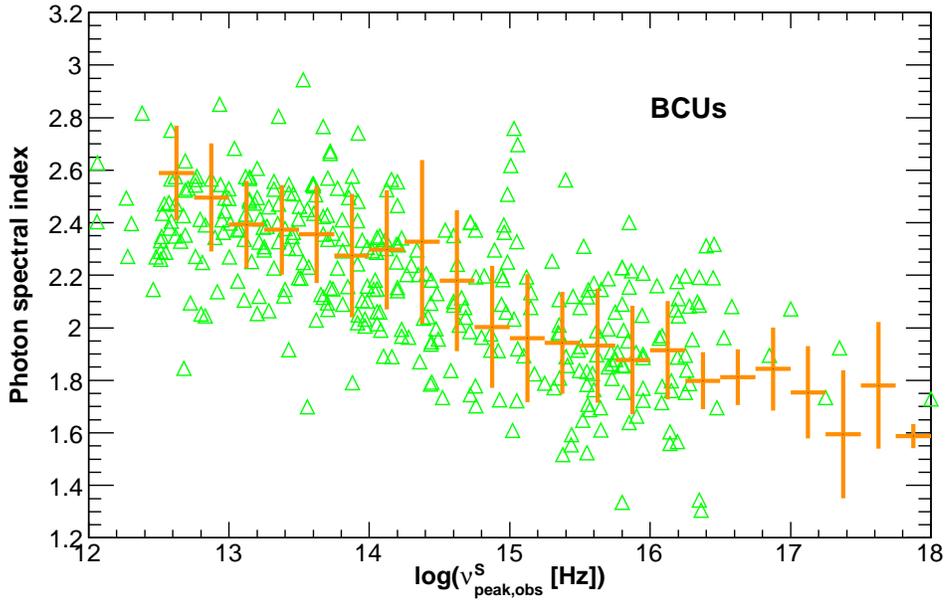}}}
\caption{Photon index versus   frequency of the synchrotron peak $\nu^S_{peak,obs}$ for blazars of unknown types (BCUs).  For comparison, the orange bars show the average index for different bins in  $\nu^S_{peak,rest}$ for blazars with known types as displayed  in Figure \ref{fig:index_nu_syn}.}
\label{fig:index_nu_syn_agu}
\end{figure}

Figure \ref{fig:index_nu_syn} shows the position of the synchrotron peak  $\nu^S_{peak,meas}$ versus the photon spectral index for FSRQs and BL~Lacs with measured redshifts. The strong anticorrelation already observed in 1LAC and 2LAC is confirmed.   
Fitting a linear function $\Gamma= A + B \log (\nu_\mathrm{peak,rest}^\mathrm{S}/10^{14} Hz)$
yields  $A=2.25\pm$0.04 and \mbox{$B=-0.18\pm0.03$}. The mean and rms of the $\Gamma$ distributions are 2.44$\pm$0.20, 2.01$\pm$0.25, 2.21$\pm$0.18, 2.07$\pm$0.20, 1.87$\pm$0.20 for FSRQs, the whole BL~Lac sample, LSP-, ISP- and HSP-BL~Lacs, respectively. FSRQs are overwhelmingly of the LSP class, so no distinction between SED-based classes will be made for them in figures and tallies. Only 37 FSRQs are of the ISP class and only 2 of the HSP class (BZB J0202+0849 and NVSS J025037+171209 associated with 3FGL J0202.3+0851 and 3FGL J0250.6+1713 respectively). As is visible in Figure \ref{fig:index_nu_syn}, most ISP-FSRQs have softer spectra than the bulk of ISP-BL~Lacs ($\langle\Gamma\rangle$=2.40$\pm$0.04 vs. 2.07$\pm$0.02). In contrast, the two HSP-FSRQs have spectra ($\langle\Gamma\rangle$=2.01) on par with the HSP-BL~Lacs and thus much harder than the spectra of most other FSRQs.  A similar trend is actually observed for BCUs as can be seen  in Figure \ref{fig:index_nu_syn_agu}, where the photon spectral index is plotted versus $\nu_\mathrm{peak,obs}^\mathrm{S}$. In this Figure, the orange bars show the average index for different bins in  $\nu^S_{peak,rest}$ obtained from the data plotted in Figure \ref{fig:index_nu_syn} for blazars of known types. This comparison supports the idea that BCUs with low  $\nu^S_{peak}$ and  high $\Gamma$ are likely FSRQs, while the rest would  mostly be BL~Lacs\footnote{ Note that Figure \ref{fig:index_nu_syn_agu}  plots  $\nu_\mathrm{peak,obs}^\mathrm{S}$ while Figure \ref{fig:index_nu_syn}  shows  $\nu_\mathrm{peak,rest}^\mathrm{S}$. The (1+z) correction, not applied in the latter case, is not expected to have a significant effect on the correlation strength.}.   

\subsection{\label{sec:z}Redshift}

\begin{figure}
\centering
\resizebox{16cm}{!}{\rotatebox[]{0}{\includegraphics{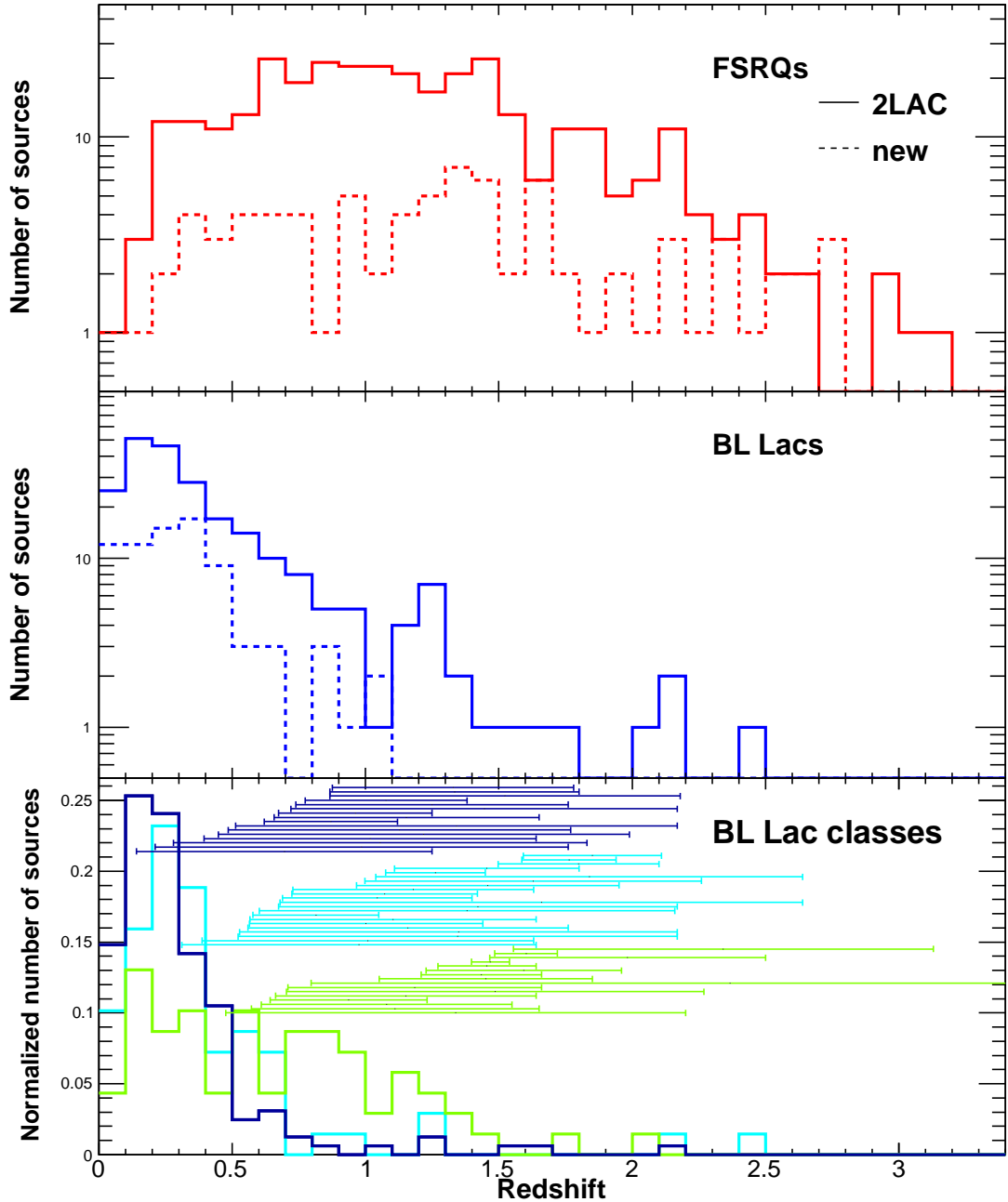}}}
\caption{Redshift distributions (solid: 2LAC sources, dashed: new 3LAC sources) for FSRQs (top), BL~Lacs (middle) and different types of BL~Lacs (bottom): LSPs (green), ISPs (light blue) and  HSPs (dark blue). The ranges between lower and upper limits are also depicted in the bottom panel when both limits are available.}
\label{fig:redshift}
\end{figure}

Figure \ref{fig:redshift} compares the redshift distributions for FSRQs and BL~Lacs in the 2LAC Clean Sample and those for the new 3LAC Clean-Sample sources (note that 50\% of the BL~Lacs do not have measured redshifts, see below). The distributions are fairly similar, although the newly detected  FSRQs are located at slightly higher redshift than the 2LAC ones ($\langle z \rangle$=1.33$\pm$0.08 vs. 1.17$\pm$0.03). The maximum redshift for an FSRQ is still 3.1 (four FSRQs have 2.94 $< z <$3.1) and has not changed since the 1LAC. This trend allowed the conclusion that the number density of FSRQs grows dramatically up to redshift $\simeq$0.5-2.0 and declines thereafter \citep{Aje12}.  

The redshift distribution of new BL~Lacs is somewhat narrower than that of the 2LAC sources, with a maximum near z=0.3. The redshift distributions gradually spread out to higher redshifts when moving from  HSP-BL Lacs to LSP-BL Lacs, a feature already seen in 2LAC. However, the HSP distribution extends to higher redshifts relative to 2LAC, with six  HSP-BL~Lacs having measured redshifts greater than 1  and one (MG4~J000800+4712) having a redshift greater than 2. Five of these  six HSPs were already included in 2LAC but either lacked measured redshifts or were classified differently.  

Among BL~Lacs, 309 have a measured redshift, while 295 do not.
The fraction of BL~Lacs without redshift is
55\%, 61\% and 40\% for LSPs, ISPs and HSPs respectively. However, \cite{Sha13} have provided  redshift constraints for 134 2LAC BL~Lacs: upper limits from the absence of Ly $\alpha$ absorption for all of them and lower limits from non-detection of the host galaxy or from intervening absorption line systems for a subset of 54 objects. It was noted by these authors that the average lower limit exceeded the average measured redshift for BL~Lacs, indicating that the measured redshifts are biased low. 
The allowed redshift ranges for the 54 sources having both lower and upper limits are plotted in the bottom panel of Figure \ref{fig:redshift} confirming that they are in tension with the measured redshift distributions, in particular for HSPs. Kolmogorov-Smirnov tests (K-S) yield probabilities of $2 \times 10^{-2}$, $1 \times 10^{-7}$, and $1 \times 10^{-6}$  that the distributions of measured redshifts and lower limits are drawn from the same underlying population for LSPs, ISPs and HSPs respectively. The redshift ranges are very similar for the different subclasses and all cluster at high redshifts, with a median around $z$=1.2. This is in good agreement with the predictions of \cite{Gio13}, which posit that most LAT-detected BL~Lacs are actually FSRQs with their emission lines swamped by the non-thermal continuum hampering determination of their redshifts.       

\subsection{Luminosity}

\begin{figure}
\centering
\resizebox{16cm}{!}{\rotatebox[]{0}{\includegraphics{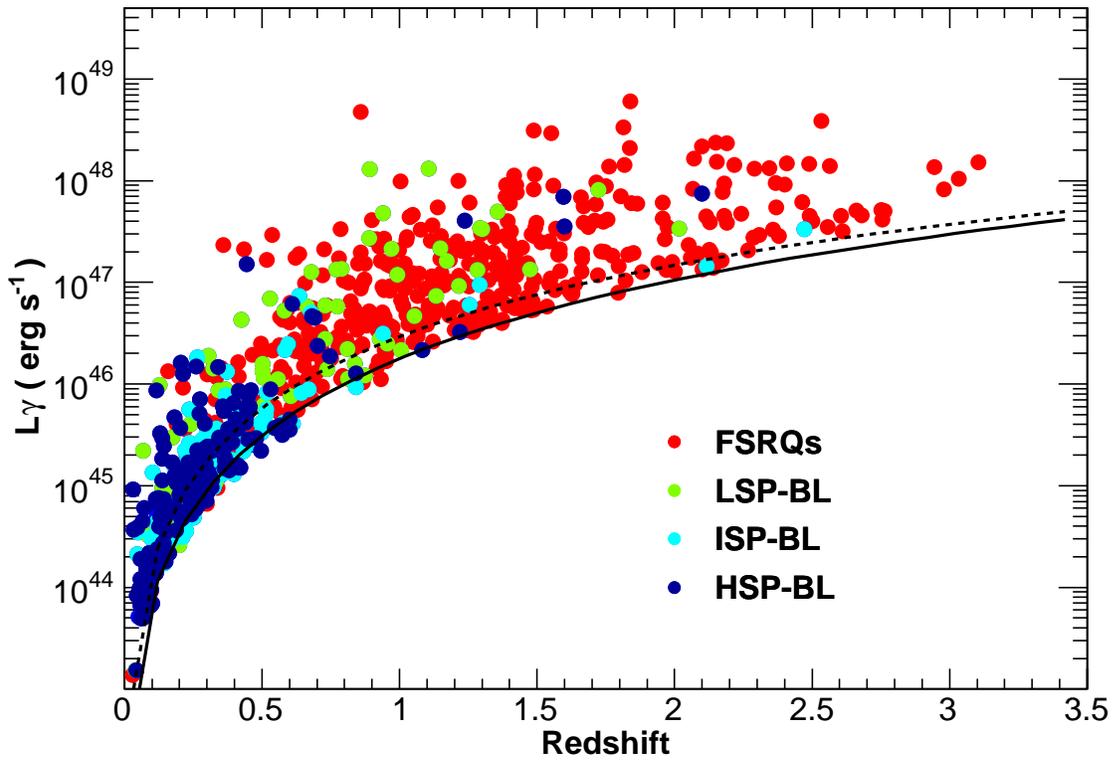}}}
\caption{Gamma-ray luminosity versus  redshift. Red: FSRQs, green: LSP-BL Lacs, light blue: ISP-BL Lacs,  dark blue: HSP-BL Lacs. The solid (dashed) curve represents the approximate detection limit for $\Gamma$=1.8 ($\Gamma$=2.2). }
\label{fig:L_redshift}
\end{figure}

The gamma-ray luminosity has been computed from the 3FGL energy flux between 100 MeV and 100 GeV, obtained by spectral fitting. Figure \ref{fig:L_redshift} displays the gamma-ray luminosity plotted against redshift, together with the sensitivity limits calculated for $\Gamma$=1.8 and 2.2.  The Malmquist bias already reported in previous catalog papers is clearly visible. Low-luminosity BL~Lacs ($<$10$^{45}$ erg s$^{-1}$) cannot be detected at $z>$ 0.4. Note that sources with luminosity greater than 5 $\times$ 10$^{47}$ erg s$^{-1}$ (64 are in 3LAC) could still be detected at $z>$ 3.2.

\begin{figure}
\centering
\resizebox{16cm}{!}{\rotatebox[]{0}{\includegraphics{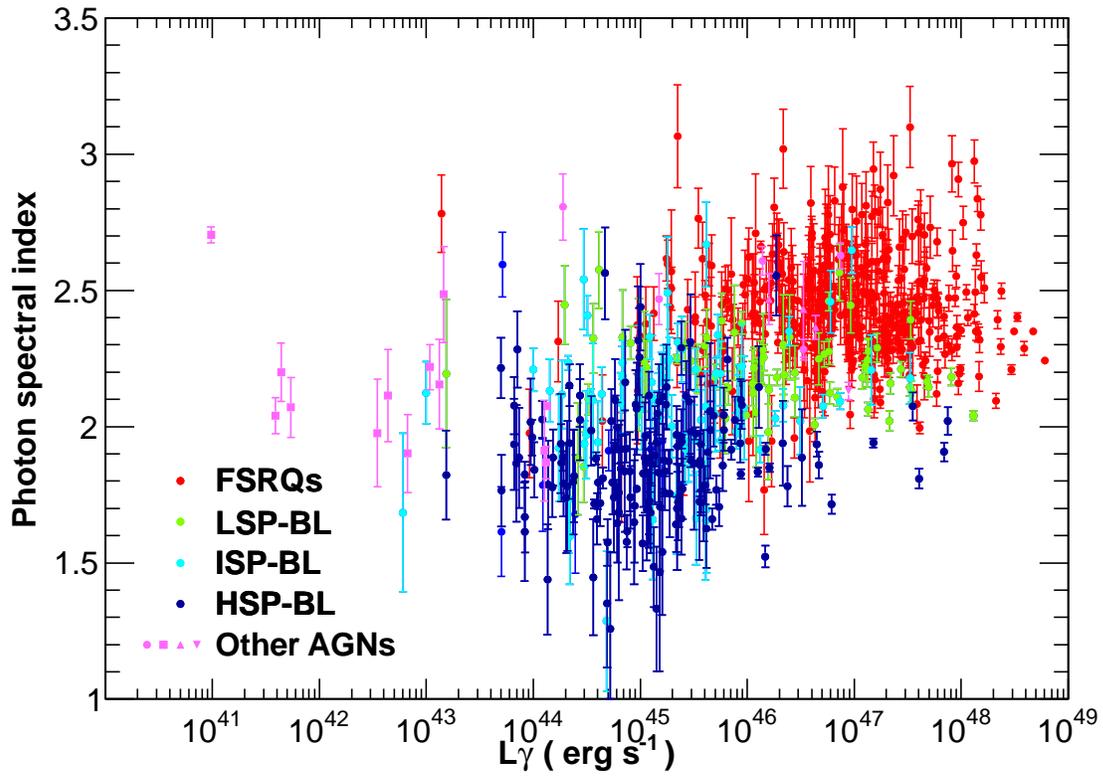}}}
\caption{Photon index versus  gamma-ray luminosity. Red: FSRQs, green: LSP-BL Lacs, light blue: ISP-BL Lacs,  dark blue: HSP-BL Lacs, magenta: other AGNs (circles: NLSy1s, squares: radio galaxies, up triangles: SSRQs, down triangles: AGNs of other types).}
\label{fig:index_L}
\end{figure}

\begin{figure}
\centering
\resizebox{16cm}{!}{\rotatebox[]{0}{\includegraphics{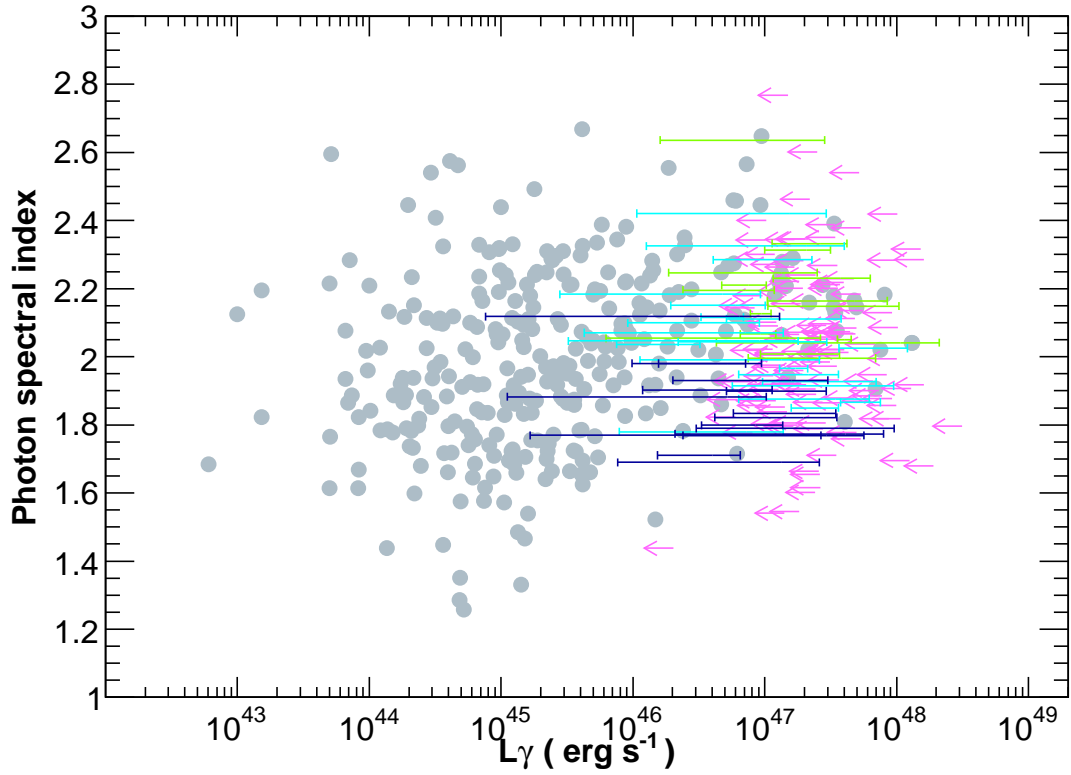}}}
\caption{Photon index versus   gamma-ray luminosity for BL Lacs. Segments are plotted for sources  having both lower- and upper-limits on their redshifts.  Green: LSP-BL Lacs, light blue: ISP-BL Lacs,  dark blue: HSP-BL Lacs. Magenta arrows are used for sources with upper limits only.    BL Lacs with measured redshifts are depicted in gray, regardless of their SED classes.}
\label{fig:index_L_lim}
\end{figure}

Figure \ref{fig:index_L} shows the LAT photon index versus the gamma-ray luminosity for the different blazar classes. This correlation has been widely discussed in the context of the ``blazar divide'' or ``blazar sequence''  \citep{Ghi09,Pad12,Mey12,Ghi12,Fin13,Gio13}. The features are similar to 2LAC, namely 
a branch of MAGNs separate from the bulk of blazars and a correlated trend of both luminosity and photon index as $\nu^S_\mathrm{peak}$ decreases. Figure \ref{fig:index_L_lim} shows the LAT photon index versus the gamma-ray luminosity for the 57 BL Lacs with both lower and upper limits on their redshifts  or only upper limits (134 sources). Because of the bias mentioned above, the HSPs with both limits are more luminous on average than those with measured redshifts, thus populating a previously scarcely occupied area in the $L_\gamma$-$\Gamma$ diagram. This observation has profound consequences for the blazar sequence.  Note that \cite{Aje14} found a small but significant correlation between gamma-ray luminosity and spectral index when including the redshift constraints from \cite{Sha13}.

\subsection{Spectral Curvature}

First observed for 3C~454.3 \citep{Abdo_3C} early in the {\it Fermi} mission, a significant curvature in the energy spectra of many bright FSRQs and some bright LSP-/ISP-BL~Lacs is now a well-established feature \citep{spec_an,1LAC}. The break energy obtained from a broken power-law fit has been found to be remarkably constant as a function of flux, at least for 3C~454.3  \citep{Abdo_3C_11}. Several explanations  have been proposed to account for this feature,  including $\gamma\gamma$ attenuation from He\,{\sc ii} line photons \citep{Pou10},
intrinsic electron spectral breaks \citep{Abdo_3C}, Ly $\alpha$ scattering
\citep{Abdo_3C_10}, Klein-Nishina effects taking place when jet electrons scatter BLR radiation in a near-equipartition approach  \citep{Cer13} and hybrid scattering \citep{Fin10}. The level of curvature has been observed to diminish during some flares \citep[e.g.,][]{Pac13}.   

\begin{figure}
\centering
\epsscale{1.35}
\plottwo{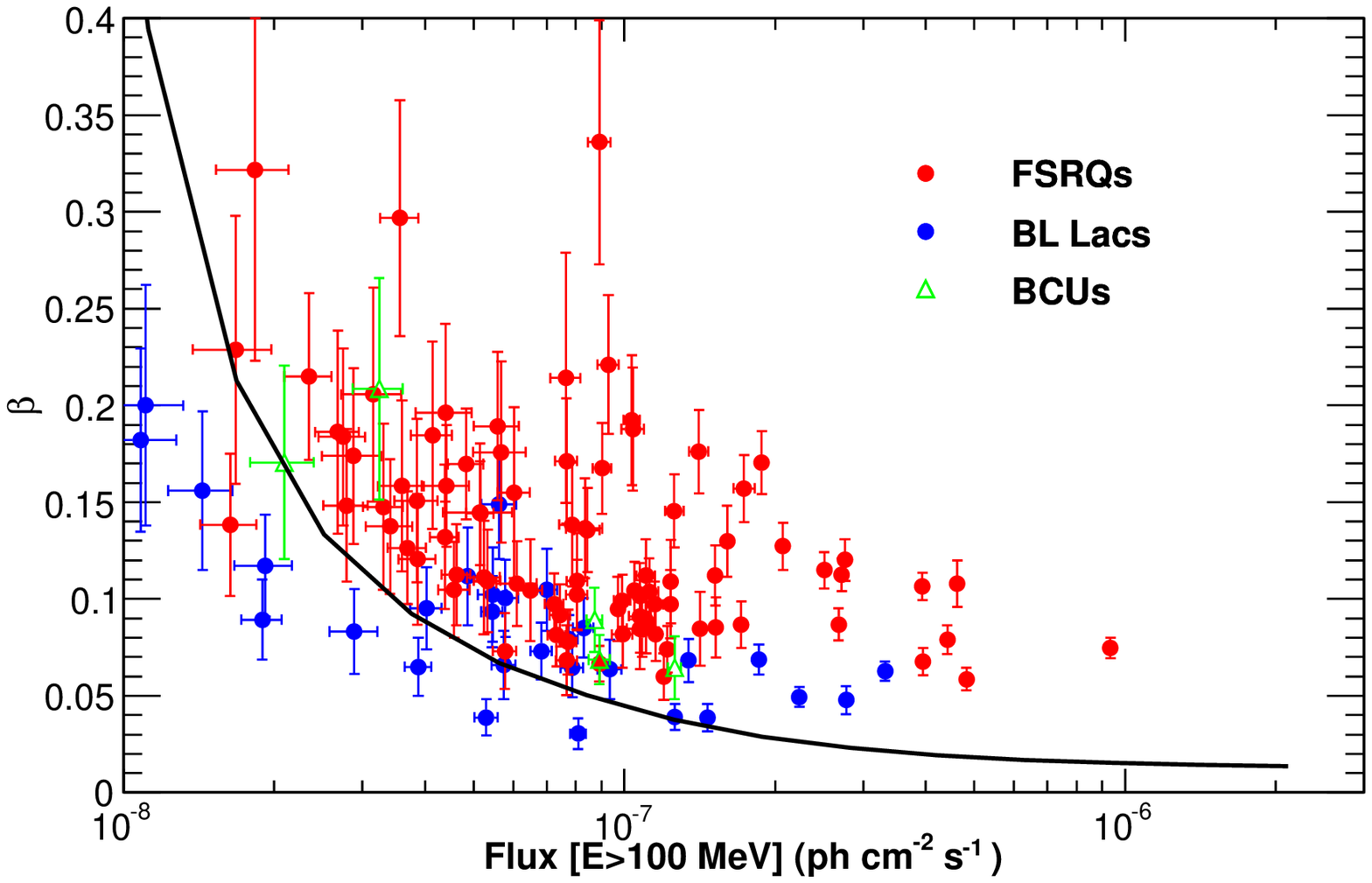}{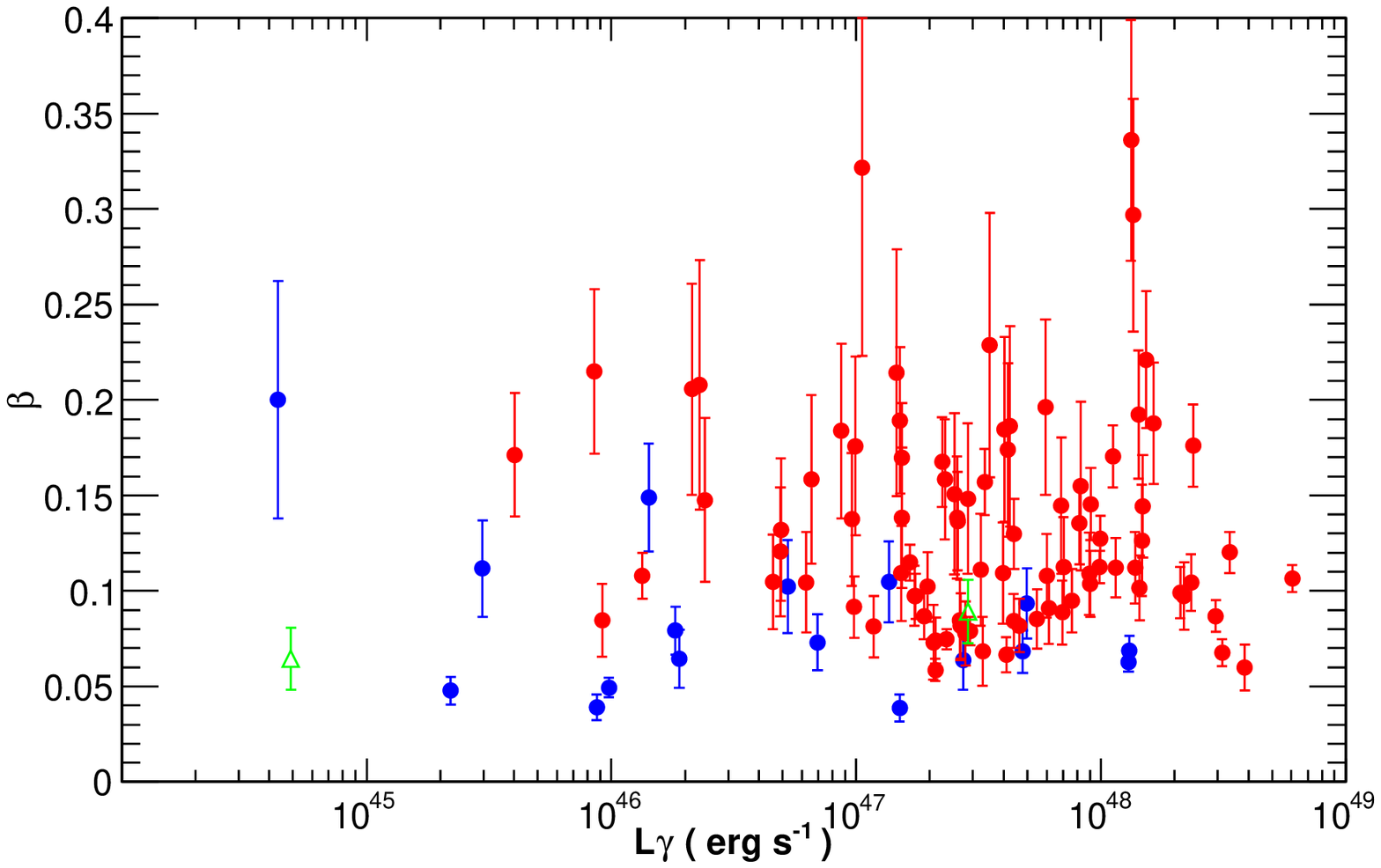}
\caption{Log-parabola parameter $\beta$ plotted versus photon flux above 100 MeV (top, the line depicts the analysis limit $TS_{curve}=16$ estimated for FSRQs) and gamma-ray luminosity (bottom). Red circles: FSRQs, blue circles: BL~Lacs, green  triangles: AGNs of unknown type.}
\label{fig:beta_flux}
\end{figure}

\begin{figure}
\centering
\resizebox{12cm}{!}{\rotatebox[]{0}{\includegraphics{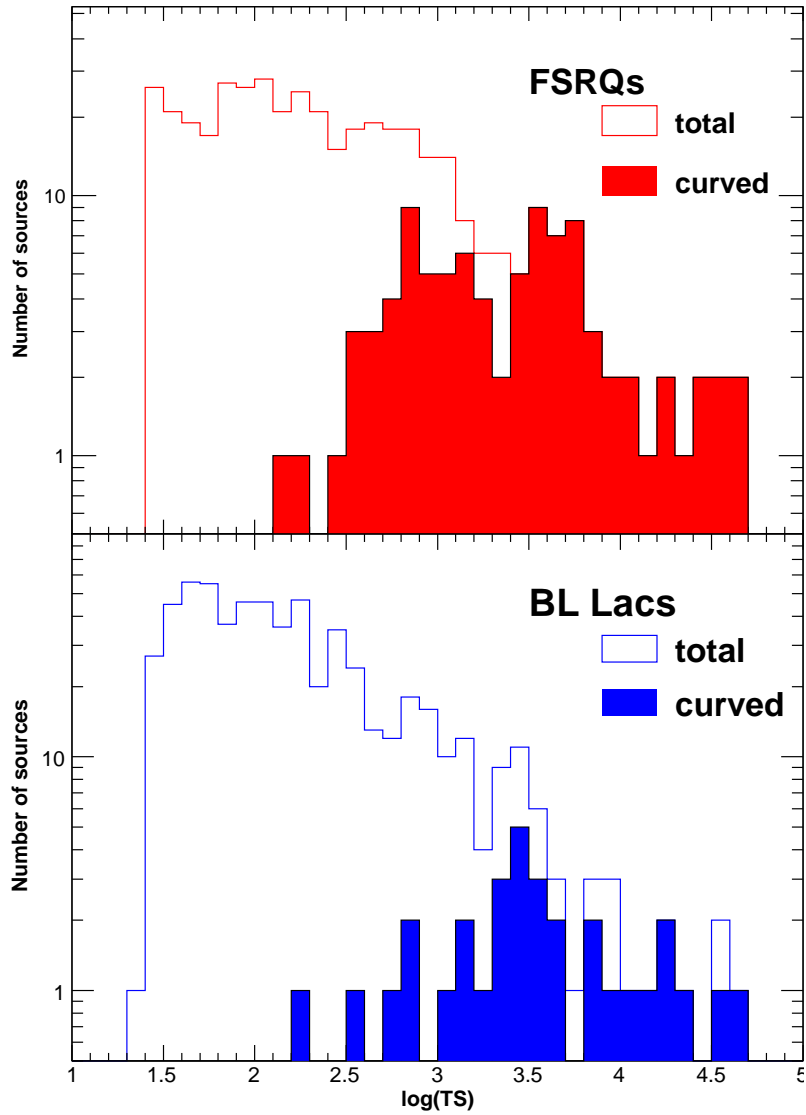}}}
\caption{$TS$ distributions of FSRQs (top) and BL~Lacs (bottom). Solid histograms: total, filled histograms: sources with significant spectral curvatures.} 
\label{fig:TS_curv}
\end{figure}

In the 3FGL analysis, a switch is made from a power-law model to a log-parabola model  whenever $TS_{curve}>$16. The spectrum of the FSRQ 3C~454.3 cannot be well fitted with a log-parabola model, a power-law+exponential cutoff being a better model.  
A total of 91 FSRQs (57 in 2LAC), 32 BL~Lacs (12 in 2LAC) and 8 BCUs show significant curvature at a confidence level $>$99\%. 
Figure \ref{fig:beta_flux} shows the log-parabola $\beta$ parameter plotted against gamma-ray flux and luminosity.  At a given flux or luminosity
the spectra of BL Lacs are less curved than those of FSRQs, a feature already reported in 2LAC.
Figure \ref{fig:TS_curv} compares the $TS$ distributions for sources with curved spectra and those for the whole samples of FSRQs and BL~Lacs. All bright FSRQs have curved spectra. For BL~Lacs, the situation is more diverse. For BL~Lac sources with $TS>$ 1000, the fraction of sources with curved  spectra is 16/23 (70\%) for  LSPs, 6/19 (32\%) for ISPs and 5/28 (18\%) for HSPs. Note that because the latter have harder spectra than LSPs/ISPs on the average, potential spectral curvature is easier to detect for them. The average $\beta$ is lower for HSPs (0.05) than for LSPs and ISPs (0.08).    

% see curv.py

\subsection{\label{sec:var} Variability}

Variability is a key feature of blazars. The 3FGL monthly averaged light curves provide a baseline reference against which other analyses can be cross-checked and enable cross-correlation studies with data obtained at other wavelengths. Although variability at essentially all time scales has been observed in blazars, the monthly binning represents a trade off between a shorter binning needed to resolve flares in bright sources and a longer binning required to detect faint sources. Even so, only 15 sources are detected in all 48 bins with monthly significance $TS>$25, while this number becomes 46 if a relaxed condition $TS>$4 is required. The 15 sources include 11 BL~Lacs (7 HSPs), only 3 FSRQs (PKS~1510$-$08, 4C~+55.17, B2~1520+31) and the radio galaxy NGC~1275. The 46 sources comprise 28 BL~Lacs (14 HSPs), 15 FSRQs, one BCU  and two radio galaxies, NGC~1275 and Centaurus A.

\begin{figure}
\centering
\resizebox{16cm}{!}{\rotatebox[]{0}{\includegraphics{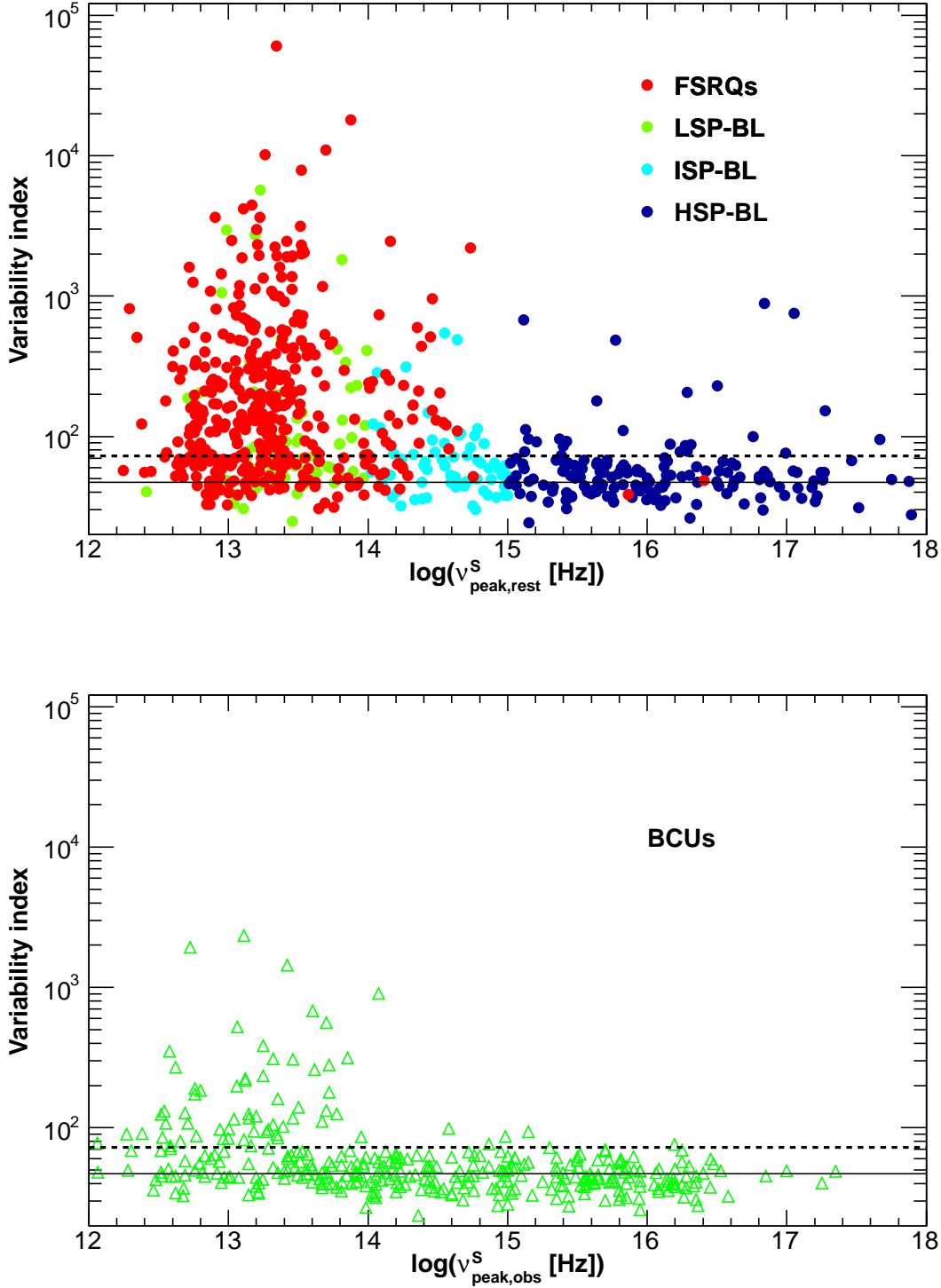}}}
\caption{Top: Variability index versus  rest-frame synchrotron peak frequency. Red: FSRQs, green: LSP-BL Lacs, light blue: ISP-BL Lacs,  dark blue: HSP-BL Lacs. The solid line depicts the average variability index expected for non-variable sources. The dashed line corresponds to the 99\% confidence level for a source to be variable. Bottom:  Variability index versus observed synchrotron peak frequency for BCUs. The lines are the same as in the upper panel.}
\label{fig:varind_sync}
\end{figure}

We will focus here on the variability index defined in Section \ref{sec:obs}; a value of 72.44 for this index indicating variability at 99\% confidence level (while the average index for non-variable sources is 47). Recall that this index can be large only  for sources that are both variable and relatively bright.  This index is plotted versus the  synchrotron peak frequency in Figure \ref{fig:varind_sync}. The features already reported in 2LAC are again visible, with a large fraction of FSRQs found to be variable (69\%), with the fraction for BL~Lacs much lower on average (23\%) and with a steadily decreasing trend as $\nu^S_{peak}$ rises  (39\%, 23\%, 15\% for LSPs, ISPs and HSPs respectively). These fractions are quite similar to those reported in 2LAC, despite the larger population and longer time span of the light curves. A similar trend between variability index and $\nu^S_{peak}$  is observed for blazars of unknown type  (Fig. \ref{fig:varind_sync} bottom) with 21\% of them found to be variable.

\begin{figure}
\centering
\resizebox{12cm}{!}{\rotatebox[]{0}{\includegraphics{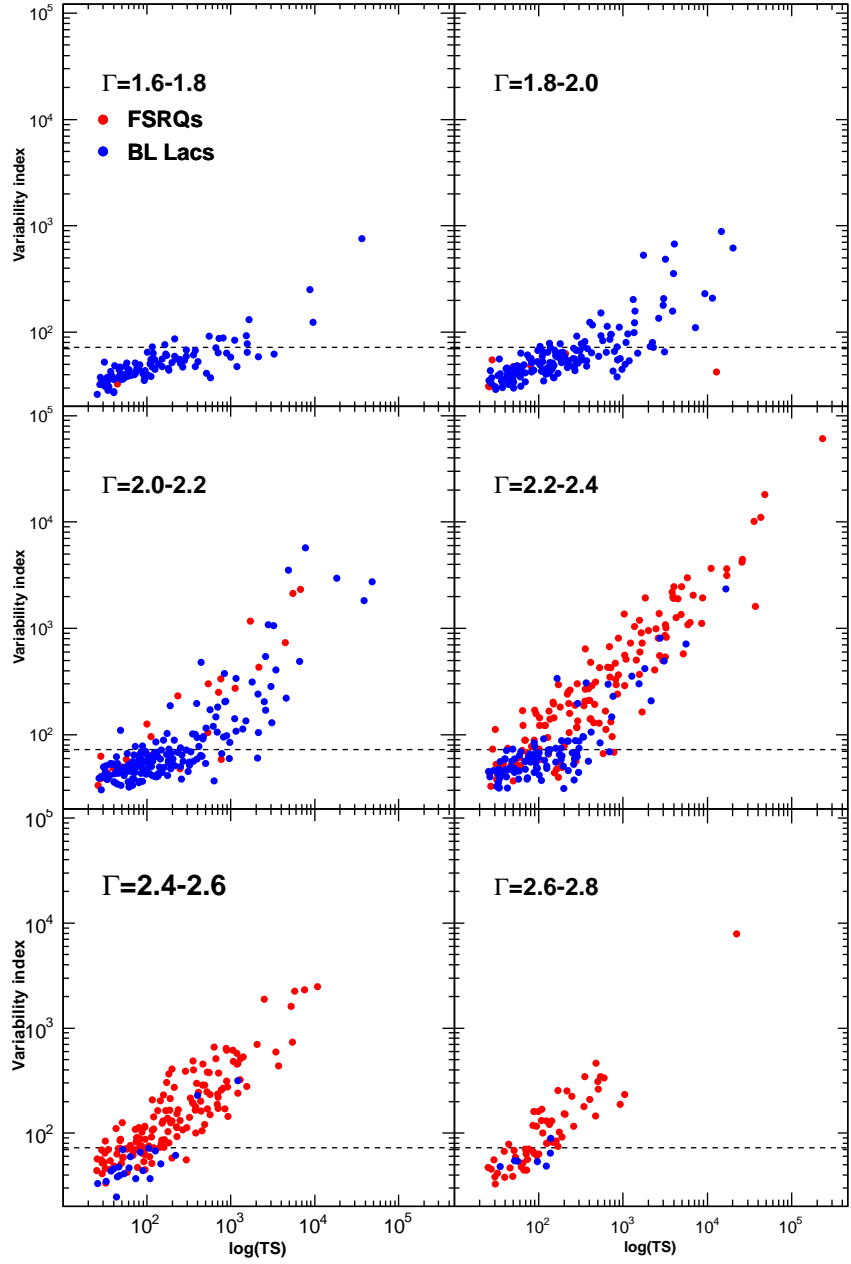}}}
\caption{Top: Variability index versus  $TS$ for 6 bins in the photon spectral index $\Gamma$. Red: FSRQs, blue: BL Lacs. 
The dashed line corresponds to the 99\% confidence level for a source to be variable.}
\label{fig:varind_TS_ind}
\end{figure}

The variability index is plotted versus TS for different bins in the photon spectral index in Figure \ref{fig:varind_TS_ind}. A distinct trend is visible: for a given $TS$ the mean variability index increases as the spectrum becomes softer (the spectral index increases) up to $\Gamma=2.4$ where this effect saturates. A net difference between FSRQs and BL Lacs is also apparent, confirming the behavior reported above. For $\Gamma>2.2$, 72\% of FSRQs and 25\% of BL~Lacs are variable above the 99\% confidence level.    

\begin{figure}
\centering
\resizebox{16cm}{!}{\rotatebox[]{0}{\includegraphics{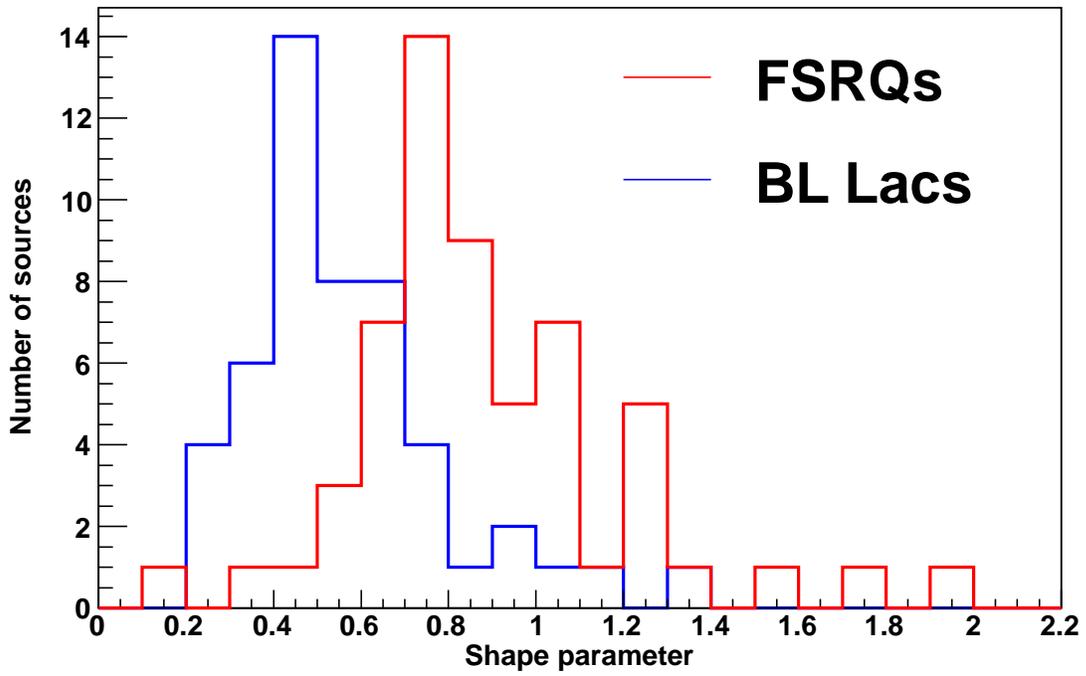}}}
\caption{Log-normal-function shape parameters $\sigma_{Ln}$ obtained from the monthly-flux distributions of $TS>1000$ FSRQs (red) and BL~Lacs (blue).}
\label{fig:lognormal}
\end{figure}

For each source, we fit the distribution of monthly photon fluxes with a lognormal function 
\begin{equation}
f_{Ln}(x) = \frac{N_{Ln}}{x \: \sigma_{Ln} \: \sqrt{2 \pi}} \: \exp \left [
- \frac{(\log(x)-u_{Ln})^2}{2\sigma_{Ln}^2} \right ],
\end{equation}
treating the flux values returned by the maximum-likelihood algorithm as if they were always significant, for simplicity. 
The  lognormal function
 has commonly been used to model blazar flux distributions \citep[e.g.,][]{Gie09,Tlu10} and provides reasonable fits for most sources of our large sample. This distribution is expected for a process involving a large number of multiplicative, independently varying parameters. 
Figure  \ref{fig:lognormal} compares the distributions of shape parameters  $\sigma_{Ln}$ of FSRQs and BL~Lacs that have been detected in 48 months above a  $TS$ of 1000 and had a monthly $TS$ above 4 in at least 24 monthly periods. These distributions are distinct. The modes are about 0.8 and 0.4 for FSRQs and BL~Lacs respectively, confirming a larger flux variability for the former.  

%{\bf Average Fractional variability about 1, was 0.5 in 2FGL}

 To further illustrate the detection variability and how the sample of brightest blazars renews itself, we compare the samples of brightest sources detected during the first and the last three-month periods of the 4 year-long data accumulation time. We applied the same TS cut used to select the LBAS sample \citep{LBAS}, namely  $TS>$100 (simply adding up the monthly $TS$ values). The two samples include similar numbers of sources (128 vs. 134), but have only 50\% (65) of the sources in common. 
 
\section{Multiwavelength Properties of 3LAC Sources}

It was shown in 2LAC that the LAT-detected blazars display on average larger
radio fluxes than non-detected blazars and that they are all bright in the optical. Tables \ref{tab:prob1} and \ref{tab:prob2} give archival data for the 3LAC and low-latitude sources respectively.  Below we focus on the connection with the two neighboring  bands, namely the hard X-rays and the VHE bands. 

\subsection{Sources Detected in Hard X-rays}

\begin{figure}
\centering
\resizebox{16cm}{!}{\rotatebox[]{0}{\includegraphics{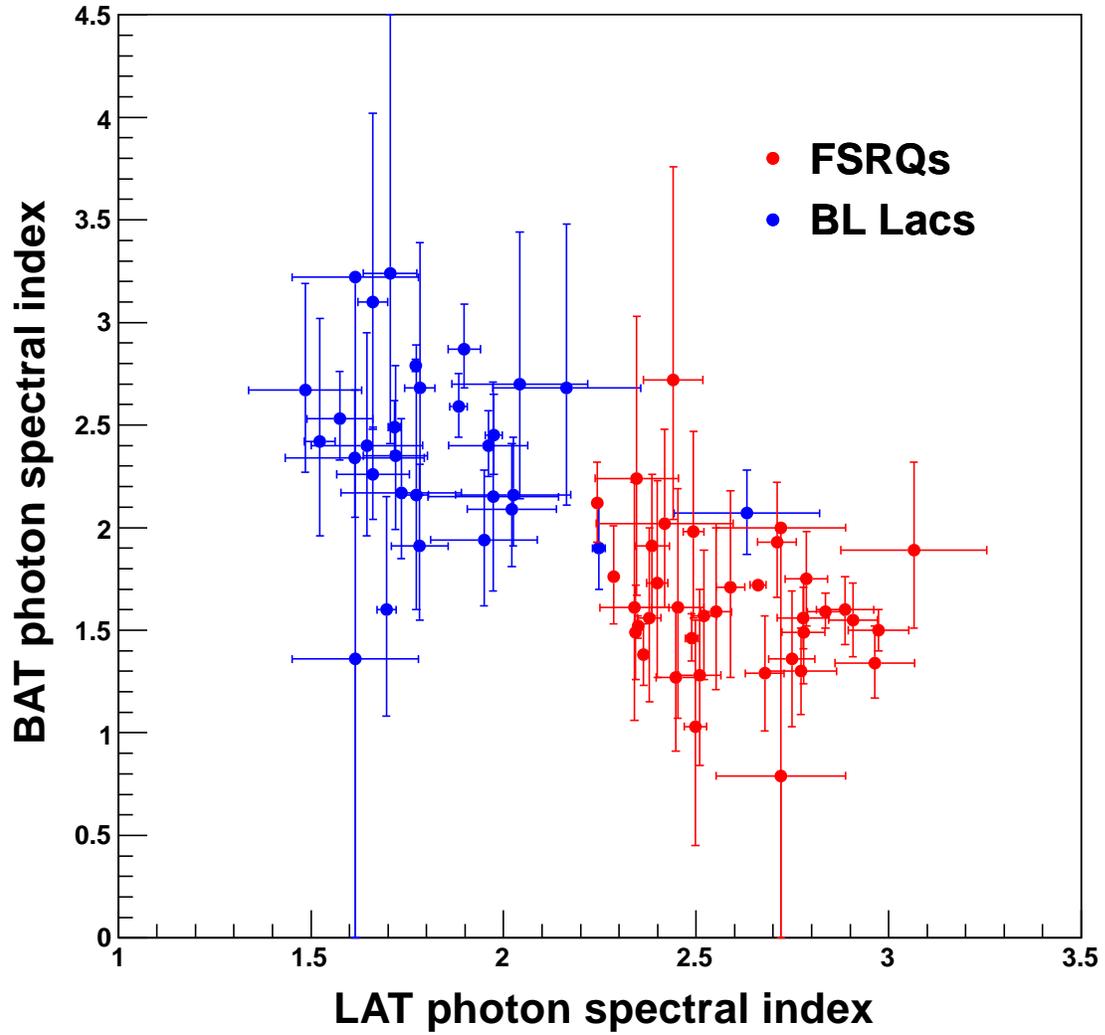}}}
\caption{Photon spectral index in the BAT band (14--195 keV) versus   photon spectral index in the LAT band.  Red: FSRQs,  blue: BL Lacs.}
\label{fig:BAT_index}
\end{figure}

A total of 85 3LAC sources are in common with the {\it Swift} BAT 70-month survey \citep{BATcatalog} in the 14--195 keV band performed between December 2004
and September 2010  (there were 47 in 2LAC). These 85 sources include 34 FSRQs with an average redshift of 1.37$\pm$0.15.  Only 9 BAT FSRQs are missing from 3LAC. The average LAT photon index of BAT-detected FSRQs is 2.57, i.e., somewhat softer than the overall average photon index of LAT FSRQs (2.43), a clue that their high-energy hump is located at slightly lower energies than the bulk of the FSRQs. Out of 37 BAT BL~Lacs, 30 have now been detected with the LAT.  These BL~Lacs  comprise 3 LSPs, 2 ISPs, and 19 HSPs, while 4 others are still unclassified. The large fraction of HSPs in this sample is not surprising as the detection of LSPs and ISPs in the hard X-ray band is hampered by their SEDs exhibiting a valley between the low- and high-energy humps in this band \citep [see ][]{Bot07}.    
Figure \ref{fig:BAT_index} displays the LAT  photon index versus the BAT photon index.  Despite large error bars in the BAT photon index and non-simultaneous measurements, a  remarkable anticorrelation (Pearson correlation factor $-$0.69), already noted in 2LAC, is observed.  For the HSP-BL~Lacs considered here, BAT probes the high-frequency (falling) part of the $\nu F_{\nu}$ synchrotron peak while the LAT probes the rising side of the inverse-Compton peak (assuming a leptonic scenario). For FSRQs, which are all LSPs in the common sample, BAT and LAT probe the rising and falling sides of the inverse-Compton peak respectively.

It is also worth noting that 96 3LAC sources (5 Radio Galaxies, 53 FSRQs, 33 BL~Lacs, 4 BCUs, 1 NLSy1) are present in the V38 {\sl INTEGRAL} source catalog\footnote{http://www.isdc.unige.ch/integral/science/catalogue} (based on 3-200 keV data taken since 2002), which includes 540 AGNs located at $|b|>10\arcdeg$. 

\subsection{Sources Detected at Very High Energies}

At the time of writing, 56 AGNs which  have been detected at TeV energies are listed in TeVCat \footnote{http://tevcat.uchicago.edu}. Among them,  55 are present in 3FGL (see Table \ref{tab:GeVTeV}), which is a remarkable result underscoring the level of synergy that has now been achieved between the high-energy and VHE domains. Only  HESS J1943+213 (a HSP BL~Lac located at b=$-1\fdg3$, affecting the possible LAT detection) is still missing from the 3FGL, but an analysis of five years of LAT data resulted in a $>1$ GeV detection \citep{pet14}. There are 15 newly detected sources  relative to 2FGL and six relative to the first {\sl Fermi}-LAT catalog of sources above 10 GeV  \citep[1FHL, ][based on 3 years of data]{1FHL}: SHBL~J001355.9$-$185406, 1ES~0229+200, 1ES~0347$-$121, 
RX~J0847.1+1133 (aka RBS 0723), MS~1221.8+2452  and 1H~1720+117. 

Not all of the 55 sources are included in the 3LAC Clean Sample, either because they are located at low Galactic latitudes or because they are flagged for different reasons. The average photon index for HSP BL~Lacs (representing 39 of the 55 AGNs) is 1.78$\pm$0.13 (rms), slightly harder than that for the whole 3LAC sample (1.88$\pm$0.22).  Only 28 out of the 55 3FGL sources are seen to be variable in the LAT energy range at a significance greater than 99\%.

\section{Discussion}

\subsection{Gamma-Ray Detected versus Non-Detected Blazars}

The   blazars detected in gamma rays after 4 years of LAT operation represent a sizeable fraction of the whole population of known blazars as listed in BZCAT.  BZCAT represents an exhaustive list of sources ever classified as blazars but is by no means complete. Although a comparison between the gamma-ray detected and non-detected blazars within that sample has no strong statistical meaning in terms of relative weights, it is nevertheless useful to look for general trends.

\begin{figure}
\centering
\resizebox{10cm}{!}{\rotatebox[]{0}{\includegraphics{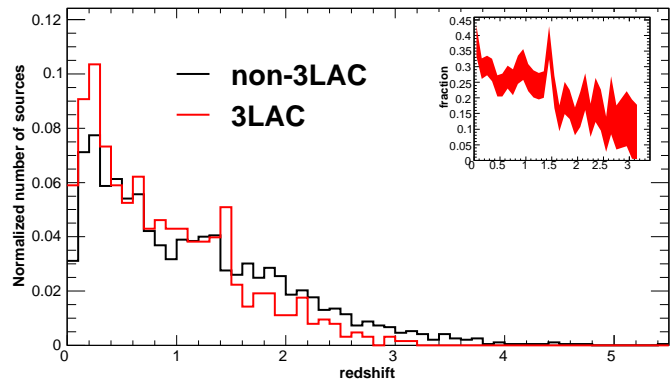}}}
\caption{ Redshift distributions for 3LAC (red) and non-3LAC  (black) BZCAT sources.  The inset shows the fraction of 3LAC sources relative to the total for a given redshift.}
\label{fig:redshift_bzcat}
\end{figure}

\begin{figure}
\centering
\resizebox{10cm}{!}{\rotatebox[]{0}{\includegraphics{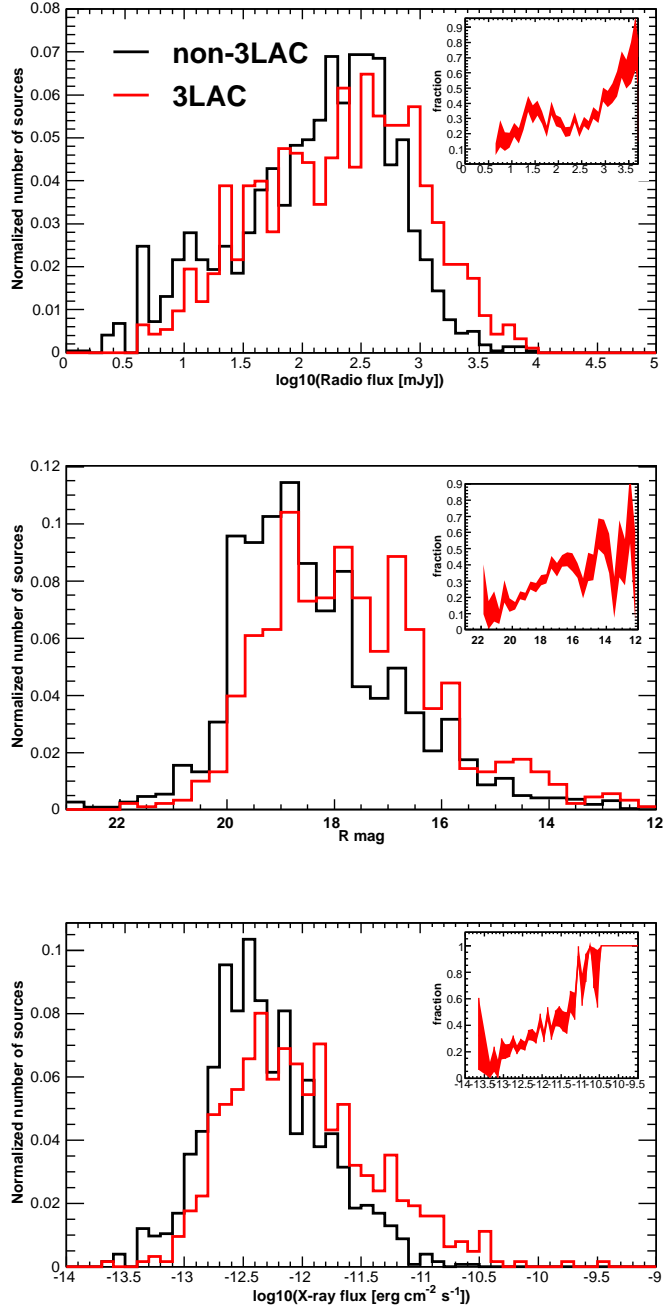}}}
\caption{ From top to bottom: radio flux density at 1.4 GHz, optical R magnitude, X-ray flux (0.1-2.4 keV) distributions for  3LAC (red) and non-3LAC (black) BZCAT sources. The insets show the fraction of 3LAC sources relative to the total for a given flux.}
\label{fig:radio_flux}
\end{figure}

\begin{figure}
\centering
\resizebox{10cm}{!}{\rotatebox[]{0}{\includegraphics{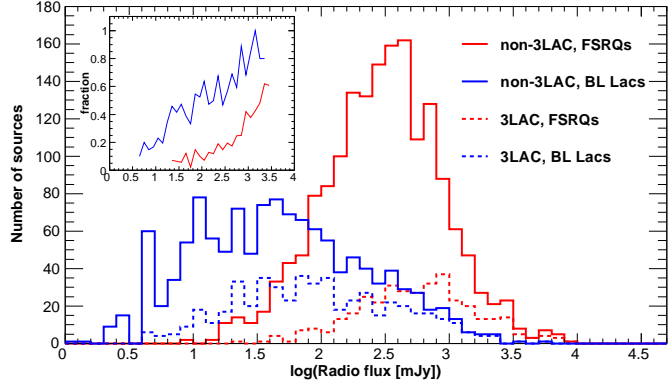}}}
\caption{Radio flux density at 1.4 GHz for 3LAC (dashed) and non-3LAC (solid) BZCAT sources. The inset displays the fraction of 3LAC sources relative to the total. Red: FSRQs, blue: BL~Lacs. }
\label{fig:radio_flux_class}
\end{figure}

The overall LAT-detected fraction is 24\% (409/1707) for FSRQs, 44\% (543/1221) for BL Lacs and  27\%  (59/221) for BCUs. A comparison between the normalized redshift distributions of the BZCAT blazars either included or not included in 3LAC  is given in  Figure \ref{fig:redshift_bzcat}, as well as the fraction of 3LAC sources relative to the total for a given redshift. A K-S test gives a probability of 3$\times 10^{-8}$ that the two redshift distributions are drawn from the same population. The distribution shapes are quite similar for the two subsets although the distribution for the blazars  undetected  by the LAT extends to significantly higher redshifts. Note that, in contrast to TeV sources, the detection of high-z sources in the LAT energy range is not strongly affected by gamma-gamma attenuation from the EBL.  The  highest-redshift BZCAT sources  (56 have z$>$3.1 reaching z=5.47) are still eluding detection by the LAT.  Figure \ref{fig:radio_flux} compares the  distributions of radio flux at 1.4 GHz, optical R-band magnitude, and X-ray (0.1-2.4 keV) flux between the BZCAT LAT-detected  and non-LAT detected blazars,  as well as the fraction of  3LAC sources relative to the total for a given flux. The gamma-ray loud blazars are somewhat brighter on average in all bands, confirming previous findings \citep{2LAC, Lis11}.   K-S tests give probabilities of  $2\times 10^{-11}$, $2 \times 10^{-22}$, and $4\times 10^{-19}$ that the 3LAC and non-3LAC distributions are drawn from the same population for the radio, optical and X-ray cases respectively. The fraction of gamma-ray loud blazars steadily decreases with  decreasing radio, optical and X-ray fluxes but remains non-negligible at the faint ends of the distributions.  
Figure \ref{fig:radio_flux_class} displays these radio-flux distributions broken down according to optical class. It is worth noting that some radio-bright blazars have not yet been detected by the LAT and that the detection fraction drops off  with decreasing radio flux in a log-linear fashion.      

\begin{figure}
\centering
\resizebox{10cm}{!}{\rotatebox[]{0}{\includegraphics{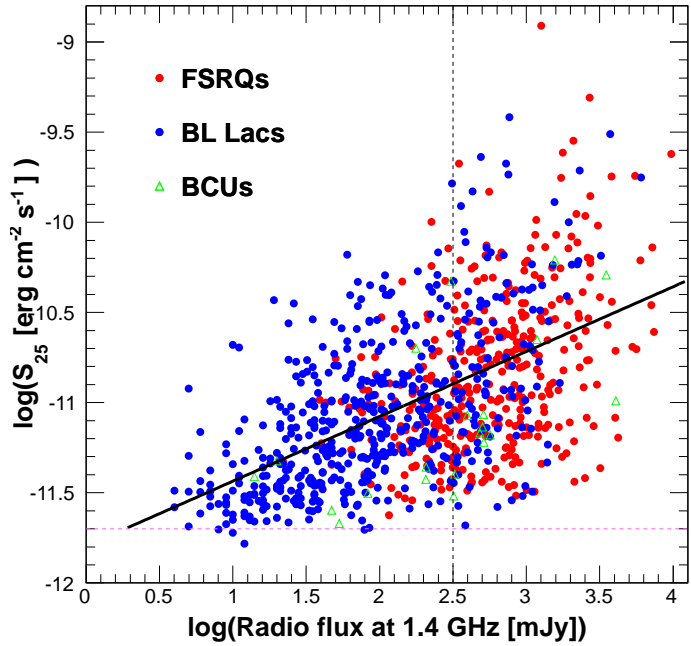}}}
\caption{Gamma-ray energy flux plotted against the radio flux density at 1.4 GHz. Red circles: FSRQs, blue circles: BL~Lacs, green triangles: BCUs. The  horizontal dashed line depicts the approximate LAT detection limit and the vertical dashed line the lower limit of the selection used in Figure \ref{fig:radio_gamma_proj}.  The solid line depicts the result of the power-law fit described in the text.}
\label{fig:radio_gamma}
\end{figure}

 In  Figure \ref{fig:radio_gamma}, the gamma-ray energy flux is plotted against the radio flux density at 1.4 GHz. A significant correlation is observed (Pearson correlation factor=0.52), confirming the findings in \citet{Ghirlanda2011,radiogamma}. The best-fit power-law relation is $F_\gamma\simeq F_r^{0.34\pm0.05}$.  Note that a stronger correlation is found if one uses the gamma-ray photon flux instead of the energy flux (Pearson correlation factor=0.72), but this results from the photon-index dependence of the flux detection threshold in the gamma-ray band already discussed above. Radio-bright FSRQs have soft spectra in the LAT band and thus high detection thresholds, reinforcing the apparent correlation between radio flux density and gamma-ray fluxes.       

\begin{figure}
\centering
\resizebox{10cm}{!}{\rotatebox[]{0}{\includegraphics{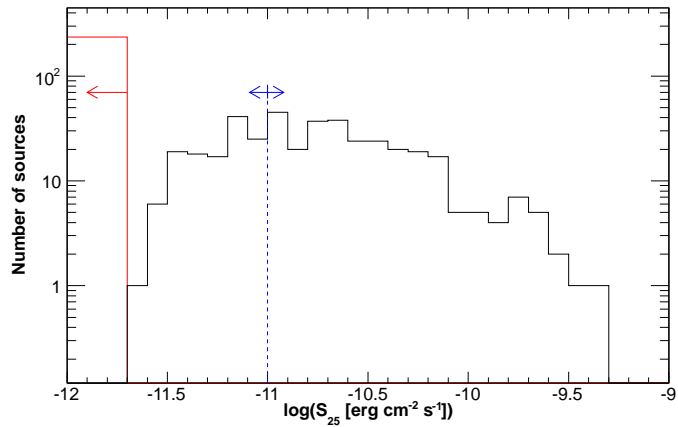}}}
\caption{Distribution of gamma-ray energy flux for LAT-detected blazars with radio flux density at 1.4 GHz above 316 mJy (black). The arrows represent the 1-$\sigma$ deviation expected for the 48-month average flux, assuming a log-normal energy-flux distribution with $\sigma_{Ln}$=1 and a mean of 10$^{-11}$ erg cm$^{-2}$ s$^{-1}$. The red upper-limit histogram schematically represents the  706 non-LAT detected BZCAT blazars with radio fluxes in the same range.}
\label{fig:radio_gamma_proj}
\end{figure}

The absence of strong difference in the redshift or flux distributions between the detected and non-detected sets of blazars supports the conjecture that they belong to the same population of sources intermittently shining in gamma rays. One can test the assumption that the fraction of non-detected sources is consistent with the variability properties assessed in Section \ref{sec:var} from the monthly light curves or if longer-term variability is required. Selecting BZCAT sources with high radio luminosity, $F_{\nu}>$316 mJy, we obtain the gamma-ray energy flux distribution plotted in Figure \ref{fig:radio_gamma_proj}. While 401 sources have been detected by the LAT,  706 sources with radio flux in the same range are not. Computing the dispersion of the 48-month flux average expected from the lognormal monthly-flux distributions presented in Section \ref{sec:var} and using the central-limit theorem, one obtains a typical value of 20\% (illustrated by the blue arrows in  Figure \ref{fig:radio_gamma_proj}
 ). This dispersion is obviously insufficient to account for the observed ratio between detected and non-detected blazars. Considerably longer time scales than those probed over the 4-year period  (associated with physical or geometrical parameter(s)  governing the observed jet gamma-ray/radio loudness ratio) must be in play. Since the fraction of LAT-detected FSRQs relative to the BZCAT total is less than that for BL~Lacs (~20\% vs. ~40\%), a larger amplitude variability of FSRQs is necessary to allow sources currently below threshold to shine in gamma rays at LAT-detection levels. This feature (a larger variability of FSRQs relative to BL~Lacs) is compatible with the observations mentioned above. 
    
\subsection{Compton Dominance}

\begin{figure}
\centering
\resizebox{9cm}{!}{\rotatebox[]{0}{\includegraphics{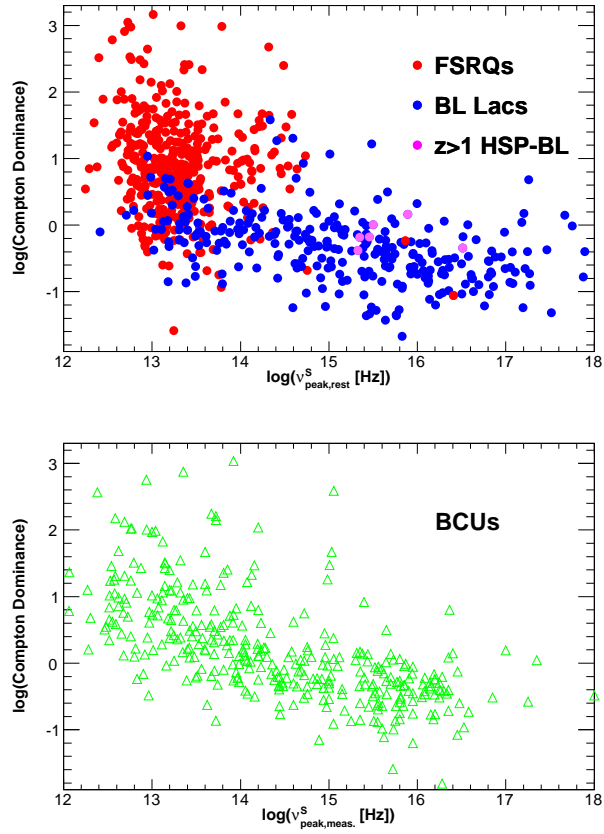}}}
\caption{Top: Compton dominance versus rest-frame peak synchrotron position. Red: FSRQs, blue: BL~Lacs, magenta: z$>$1 HSP-BL~Lacs. Bottom:  Compton dominance versus observer-frame peak synchrotron position for BCUs.}
\label{fig:cd}
\end{figure}

We consider here  the Compton dominance ratio ($CD$), i.e., the ratio between the peak $\nu F_\nu$ for the high- and low-frequency SED humps,  computed as described in \cite{SEDpaper} and \cite{Fin13}. The top panel of Figure \ref{fig:cd} shows this ratio plotted against  $\nu^S_{peak}$  \citep[similar to Figure 5 in ][ using  2LAC data]{Fin13}.  It is found that $\log CD$ has (mean, rms) of (0.60, 0.65) for FSRQs and  (-0.11, 0.48) for  LSP-BL~Lacs, while it has  (-0.39, 0.42) for ISP-BL~Lacs, and (-0.78, 0.39) for HSP-BL~Lacs.           

The spread in $CD$ is partially driven by variability. The SED data are not simultaneous, especially for FSRQs, as some of them have displayed flux variations in gamma rays greater than two orders of magnitude during the {\it Fermi} mission. However as shown in Figure \ref{fig:radio_gamma_proj}, the overall effect of variability on the mean gamma-ray flux is quite limited (see more below). 

  The combination of different beaming factors for the two humps \citep[as expected if inverse-Compton off an external radiation field is important, e.g., in FSRQs,][]{Der95}  and  different jet angles relative to the line of sight within the 3LAC sample is likely to add to this spread.   FSRQs have on average higher Compton dominance than BL~Lacs, which exhibit a trend to lower $CD$ values  with increasing $\nu^S_{peak}$. Interestingly, as can be seen from  Figure \ref{fig:cd},  the six luminous HSP-BL~Lacs located at redshift greater than 1  show $CD$ values very similar to those located at low redshift. These objects have a mean photon index of 1.94, comparable to the mean value of the whole HSP sample (1.88). Together, these features indicate that the overall SED shape of HSP-BL~Lacs  is not strongly dependent on redshift and thus neither on luminosity. 

The lower panel of Figure \ref{fig:cd} shows the corresponding plot for BCUs. Although $\nu^S_\mathrm{peak}$ has not been corrected by $(1+z)$ for most sources as their redshifts are unknown, the observed trend is very similar to that of blazars with known types. 

\begin{figure}
\centering
\resizebox{9cm}{!}{\rotatebox[]{0}{\includegraphics{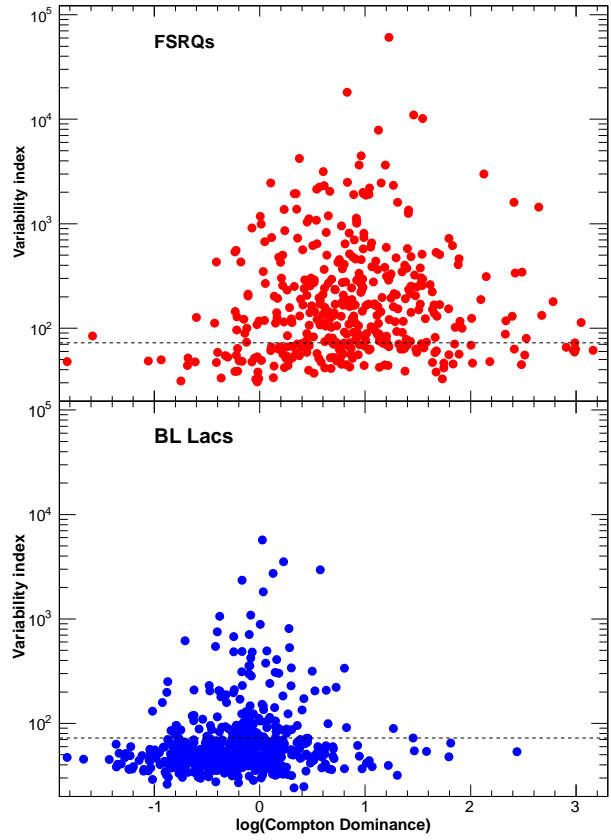}}}
\caption{Variability index versus  Compton dominance. Top: FSRQs, bottom: BL~Lacs. The dashed line corresponds to the 99\% confidence level for a source to be variable.}
\label{fig:varind_cd}
\end{figure}

An  interesting point regards the comparison between LSP-BL~Lacs and FSRQs. The gamma-ray properties of the former being intermediate between those of FSRQs and of HSP-BL~Lacs, they could be FSRQs ``in disguise'' where the emission lines are swamped by a strong non-thermal continuum as suggested by \cite{Gio13}.  Figure \ref{fig:varind_cd} shows the variability index plotted against $CD$ for FSRQs and BL~Lacs. It is seen that the regions occupied by the BL~Lacs and FSRQs have moderate overlap. %This restricts the location of transitional objects between the two classes.   
 
\subsection{log~N-log~S}

\begin{figure}
\centering
\resizebox{9cm}{!}{\rotatebox[]{0}{\includegraphics{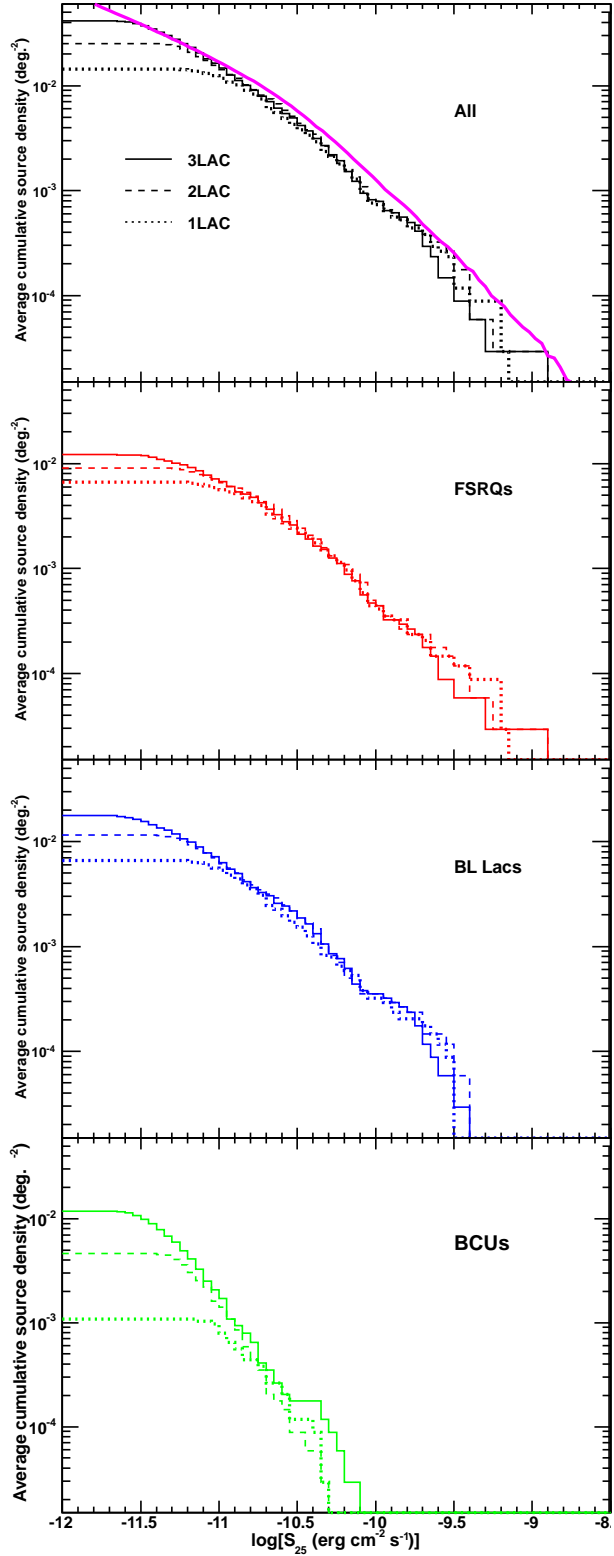}}}
\caption{Cumulative energy flux distributions (uncorrected for non-uniform sensitivity and detection/association efficiency) for blazars in Clean Samples. Solid: 3LAC, dashed: 2LAC, dotted: 1LAC. Top: Total. The magenta curve corresponds to the predictions derived from \cite{igrb}. Second: FSRQs. Third: BL~Lacs. Bottom: blazars of unknown type.  }
\label{fig:logn_logs}
\end{figure}

Figure \ref{fig:logn_logs} shows the log~N-log~S  (S being the gamma-ray energy flux and N the cumulative number of sources above this flux) plot for the full 1LAC, 2LAC, 3LAC catalogs, as well as for FSRQs, BL~Lacs and BCUs in the respective Clean Samples, uncorrected for coverage. Note that the LAT limiting energy flux is essentially independent of the photon index and thus of the blazar class as illustrated in Figure \ref{fig:index_S}.  A steady increase in the number of sources is observed for all classes, with the 3LAC being roughly in line with  extrapolations from the 2LAC. Power-law fits performed on the 3LAC distributions between somewhat arbitrary energy-flux limits (see Figure \ref{fig:logn_logs}) yield slopes of 1.23, 1.22 and 1.09 for the whole set, FSRQs and BL~Lacs respectively. Integrating the energy-flux distributions above  100 MeV  in the range  10$^{-11}$-10$^{-9}$ erg\,cm$^{-2}$\,s$^{-1}$ gives gamma-ray intensities for all sources and FSRQs of
1.4$\times$10$^{-6}$ GeV\,cm$^{-2}$\,s$^{-1}$\,sr$^{-1}$ and 4.7$\times$10$^{-7}$ GeV\,cm$^{-2}$\,s$^{-1}$\,sr$^{-1}$, respectively. These results can be compared to those obtained in assessing the diffuse gamma-ray emission \citep{igrb} : the intensity for all resolved sources at
$|b|>$20$\arcdeg$ is estimated to be 9.5$\times$10$^{-7}$ GeV\,cm$^{-2}$\,s$^{-1}$\,sr$^{-1}$. This corresponds to  1.2$\times$10$^{-6}$ GeV\,cm$^{-2}$\,s$^{-1}$\,sr$^{-1}$ after applying the geometrical correction (from $|b|>20\arcdeg$ to $|b|>10\arcdeg$), in reasonable agreement with the 3LAC-based estimate.

\section{Conclusions}

We have presented the third catalog of LAT-detected AGNs (3LAC), based on 48 months of LAT data.  This is an improvement over the 1LAC (11 months of data) and 2LAC (24 months of data) also in terms of data quality and analysis methods. Key results from the 3LAC sample include: 

1. An increase of 71\% in the number of blazars relative to 2LAC stems from the two-fold increase in exposure and the use of improved counterpart catalogs. The energy-flux distributions of the different blazar populations are in good agreement with extrapolation from earlier catalogs.

2. A significant increase of the non-blazar population  is found with respect to previous catalogs. The new sources include: two FRIIs (Pictor A, 3C~303), three FRIs (4C~+39.12, 3C~189, 3C~264 plus one possible association, Fornax A) and four SSRQs (TXS~0826+091, 4C~+0.40, 3C~275.1, 3C~286). However other sources (3C~407, NGC~6951,  NGC~6814) reported in previous catalogs are now missing.
 
3. A large fraction ($>75$\%) of {\sl Swift} hard X-ray BAT-detected blazars and all but one TeV-detected AGNs have now been detected by the ${\it Fermi}$-LAT.  

4. The most distant 3LAC blazar is the same as in 1LAC and 2LAC,  PKS~0537$-$286 lying  at z=3.1. Many BZCAT blazars at higher redshifts have yet to be detected by the LAT.  Although 50\% of the BL~Lacs still do not have measured redshifts, upper limits have recently been obtained for 134 2LAC sources and lower limits as well for 57 of them. These constraints indicate that the measured redshifts are biased low for BL~Lacs. Using the luminosities derived from these constraints, the sources populate a previously scarcely occupied area in the $L_\gamma$-$\Gamma$ diagram, somewhat undermining the picture of the blazar sequence. 

5. Along the same lines, a few rare outliers (four high-luminosity HSP BL~Lacs and two HSP FSRQs) are included in the 3LAC, while they were missing in 2LAC.   The high-luminosity HSP-BL~Lacs exhibit  Compton dominance values similar to the bulk of that class.  
 
6. The main properties of blazars previously reported in 1LAC and 2LAC are confirmed. The average photon index, gamma-ray luminosity, flux variability, spectral curvature monotonically evolve from FSRQs to HSP BL Lacs with  LSP- and ISP-BL~Lacs showing intermediate behavior.  %Overall, LSP-BL Lacs match the picture of FSRQs with beamed emissions overpowering broad line radiation.

7. The fraction of 3LAC blazars in the total population of blazars listed in BZCAT remains non-negligible even at the faint ends of the BZCAT-blazar radio, optical and X-ray flux distributions, which is a clue that even the faintest, and thus possibly all,  known blazars could eventually shine in gamma rays at LAT-detection levels.   A larger fraction (44\%) of the known BL~Lacs than FSRQs (24\%) has been detected so far.  The duty cycle of FSRQs appears to be longer than four years if most of them are eventual gamma-ray emitters.

The 3LAC catalog is intended to serve as a valuable resource for a better understanding of the gamma-ray loud AGNs. The next LAT AGN catalog will benefit from the improved Pass 8 data selection and IRFs \citep{Atw13}. Pass 8 is the result of a comprehensive revision of the entire event-level analysis,  based on the experience gained in the prime phase of the mission.  The gain in effective area  at the low end of the LAT energy range will be particularly notable. The 4LAC catalog is thus  expected to include a non-incremental number of new, especially soft-spectrum AGNs.     

\section{Acknowledgments}

\acknowledgments The \textit{Fermi} LAT Collaboration acknowledges generous
ongoing support from a number of agencies and institutes that have supported
both the development and the operation of the LAT as well as scientific data
analysis.  These include the National Aeronautics and Space Administration and
the Department of Energy in the United States, the Commissariat \`a l'Energie
Atomique and the Centre National de la Recherche Scientifique / Institut
National de Physique Nucl\'eaire et de Physique des Particules in France, the
Agenzia Spaziale Italiana and the Istituto Nazionale di Fisica Nucleare in
Italy, the Ministry of Education, Culture, Sports, Science and Technology
(MEXT), High Energy Accelerator Research Organization (KEK) and Japan
Aerospace Exploration Agency (JAXA) in Japan, and the K.~A.~Wallenberg
Foundation, the Swedish Research Council and the Swedish National Space Board
in Sweden. Additional support for science analysis during the operations phase is gratefully acknowledged from the Istituto Nazionale di Astrofisica in Italy and the Centre National d'\'Etudes Spatiales in France.

This research has made use of data obtained from the high-energy Astrophysics Science Archive
Research Center (HEASARC) provided by NASA's Goddard
Space Flight Center; the SIMBAD database operated at CDS,
Strasbourg, France; the NASA/IPAC Extragalactic Database
(NED) operated by the Jet Propulsion Laboratory, California
Institute of Technology, under contract with the National Aeronautics and Space Administration. This research has made use of data archives, catalogs and software tools from the ASDC, a facility managed by the Italian Space Agency (ASI).
Part of this work is based on the NVSS (NRAO VLA Sky Survey).
The National Radio Astronomy Observatory is operated by Associated Universities, Inc., under contract with the National Science Foundation.
This publication makes use of data products from the Two Micron All Sky Survey, which is a joint project of the University of
Massachusetts and the Infrared Processing and Analysis Center/California Institute of Technology, funded by the National Aeronautics
and Space Administration and the National Science Foundation.
This publication makes use of data products from the Wide-field Infrared Survey Explorer, which is a joint project of the University of California, Los Angeles, and
the Jet Propulsion Laboratory/California Institute of Technology,
funded by the National Aeronautics and Space Administration.
Funding for the SDSS and SDSS-II has been provided by the Alfred P. Sloan Foundation,
the Participating Institutions, the National Science Foundation, the U.S. Department of Energy,
the National Aeronautics and Space Administration, the Japanese Monbukagakusho,
the Max Planck Society, and the Higher Education Funding Council for England.
The SDSS Web Site is http://www.sdss.org/.
The SDSS is managed by the Astrophysical Research Consortium for the Participating Institutions.
The Participating Institutions are the American Museum of Natural History,
Astrophysical Institute Potsdam, University of Basel, University of Cambridge,
Case Western Reserve University, University of Chicago, Drexel University,
Fermilab, the Institute for Advanced Study, the Japan Participation Group,
Johns Hopkins University, the Joint Institute for Nuclear Astrophysics,
the Kavli Institute for Particle Astrophysics and Cosmology, the Korean Scientist Group,
the Chinese Academy of Sciences (LAMOST), Los Alamos National Laboratory,
the Max-Planck-Institute for Astronomy (MPIA), the Max-Planck-Institute for Astrophysics (MPA),
New Mexico State University, Ohio State University, University of Pittsburgh,
University of Portsmouth, Princeton University, the United States Naval Observatory,
and the University of Washington.

{\it Facilities:} \facility{{\it Fermi} LAT}.

\clearpage
\begin{appendix}

%%%%%%%%%%%%%%%%%%%%%%%%%%%%%%%%%%%%%%%%%%%%%%%%%%%%%%%%%%%%%%%%%%%%%%%%%%%%%%%%%%%%%%%
\section{Note on convention for source association counterpart nomenclature}
%%%%%%%%%%%%%%%%%%%%%%%%%%%%%%%%%%%%%%%%%%%%%%%%%%%%%%%%%%%%%%%%%%%%%%%%%%%%%%%%%%%%%%%

In this paper we have tentatively adopted a  history-based rationale for the names of blazar and other AGN source counterparts associated with 3LAC sources, as reported in the 3FGL catalog FITS file \footnote{http://fermi.gsfc.nasa.gov/ssc/data/access/lat/4yr\_catalog/gll\_psc\_v16.fit}. This naming rationale is already working as the source name resolver in NED (NASA/IPAC Extragalactic Database), and was already in use, in part, in the 2LAC paper. It is possible to retrieve an approximate knowledge about the chronological appearance of a radio/optical/X-ray point source in past catalogs thanks to NED, Simbad-Vizier and ADS databases. The best-known (widely-used) naming rationale is more arbitrary and more difficult to reconstruct, it suffers more from subjectivity, and  applies only to the minority of the brightest blazars/AGNs.

AGNs and blazars were first discovered as optical non-star-like/nebula objects (i.e. galaxies, M, NGC, IC catalogs published between 1781 and 1905), as optical variable stars (Argelander designations for BL Lac, W Com, AP Lib), unusually optically blue starlike objects (Ton, PHL, Mkn catalogs all published between about 1957 and 1974), subsequent catalogs of normal or peculiar galaxies (CGCG, MCG, CGPG, UGC, Ark, Zw/I-V, Tol catalogs all published between about 1961 and 1976). Subsequent optical catalogs like the PG, PB, US, SBS, PGC, LEDA, HS, SDSS are also used in our 3LAC associations naming rationale\footnote{For all catalogs cited in this appendix, the pertaining literature and bibliographic references can be directly retrieved through the NED web database at \texttt{ned.ipac.caltech.edu/cgi-bin/catdef?prefix=XYZ}, where "\texttt{XYZ}" is the catalog/list code or prefix (e.g., "B2").}. In parallel, most blazars and AGNs were detected as new discrete point sources in the first radio observations and surveys (sources like Vir A, Cen A, Cen B, Per A etc. in early  1950s, then the 3C, CTA, PKS, 4C, O[+letter], VRO, NRAO, AO, DA, B2, GC, S1/S2/S3 catalogs all published between about 1959 and 1974). Other subsequent radio catalogs like the TXS, 5C, S4/S5, MRC, B3 (all about 1974-1985) and MG1/MG2/MG4, 87GB, 6C/7C, JVAS, PMN, EF, CJ2, FIRST, Cul, GB6, FBQS, WN, NVSS, CLASS, IERS, SUMSS, CRATES (all after 1986) are also used in our work. Other catalogs of interest at IR or UV frequencies for purposes of 3LAC association names are the KUV, EUVE, 2MASSi, 2MASS. Additional blazars that are fainter in the radio/optical bands were discovered directly thanks to the first X-ray observations (2A, 4U, XRS, EXO, H/1H, MS, 1E, 1ES, 2E, RX all published from about 1978 to mid 1990s). The subsequent (after 1997) reanalysis and catalog constructions based mainly on the {\sl ROSAT} survey and radio-X-ray source cross correlations are also used in the 3LAC (RGB, RBS, RHS, 1RXS, XSS catalogs).

The most common source counterpart roots in 3LAC associations have origin in the 3C, 4C, PKS, O[+letter], B2, S2/S3/S5, TXS, MG1/MG2, PMN, GB6, SDSS, 1ES, RX, RBS, 1RXS catalogs. PKS (Parkes Radio Catalog, Australia) chronologically is the source name preferred for southern celestial radio sources, over almost all the other epoch-overlapping radio catalogs. Likely the survey for northern celestial radio sources at Parkes started after the more  easily observable southern sources, therefore later than the O[+letter] (Ohio State University Radio Survey Catalog, USA) observations, and certainly after the 3C and 4C catalogs. The procedure for selecting source counterpart names is tuned 
to the most-used/known criterion for the most famous sources (for example OJ 287 instead of PKS 0851+202/ PG 0851+202, but PKS 0735+17 instead of OI 158 / DA 237). Other famous blazars/AGN sources are more likely to follow the best-known criterion (example: Cen A is more frequently used than NGC 5128, even though this galaxy was first discovered in the NGC catalog). For the northern celestial hemisphere the preferred radio source name chosen following the approximate chronological criterion  follows the sequence of radio catalogs reported above  (3C, CTA, 4C, O[+letter], NRAO, AO, DA, B2, GC, S1/S2/S3, TXS, MG1/MG2/MG4 etc.). Some catalog designations (like the 87GB and rare optical names) are essentially not used in the 3LAC.  RBG names have been preferred to RBS and 1RXS names, and the NVSS names have  been preferred to the SDSS names. We do not have a preference between GB6 and RX names or between RBS and 1RXS names, all being used arbitrarily.
\end{appendix}
\clearpage

\clearpage

\clearpage


%\begin{figure}
%\centering
%\resizebox{16cm}{!}{\rotatebox[]{0}{\includegraphics{index_lum_syn.eps}}}
%\caption{Photon index as a function of  synchrotron luminosity for FSRQs (top) and BL Lacs (bottom). The lines correspond to the results of linear fits.}
%\label{fig:index_lum_syn}
%\end{figure}

\clearpage

%\begin{figure}
%\centering
%\resizebox{16cm}{!}{\rotatebox[]{0}{\includegraphics{L_varind.eps}}}
%\caption{Variability index as a function of  gamma-ray luminosity for FSRQ, blue: BL~Lacs. The solid line depicts the average variability index expected for non-variable sources. The dashed line corresponds to the 99\% confidence level for a source to be variable.}
%\label{fig:L_varind}
%\end{figure}

\clearpage 


\begin{deluxetable}{|l|cc|ccc|ccc|}
\tablecolumns{9}
\tabletypesize{\normalsize}
\tablecaption{Comparison of association methods  in terms of total number of associations, estimated number of false associations ($N_{false}$),  and number of sources  associated only via a given method, $N_S$.\label{tab:assoc}} 
\tablewidth{0pt}
\tablehead{ 
\colhead{Sample}    & \multicolumn{2}{|c|}{All Methods }               & \multicolumn{3}{|c|}{ Bayesian Method} & \multicolumn{3}{|c|}{ LR Method}\\
 \colhead{}    &  \colhead{Total} & \colhead{$N_{false}$} &    \colhead{Total} &  \colhead{$N_{false}$} &  \colhead{N$_S$ }  &   \colhead{Total} & \colhead{$N_{false}$} &  \colhead{N$_S$}  
}
\startdata
 All         & 1591 & 29.7 & 1529 & 34.5 & 379 & 1212 & 120.5 & 62 \\
 Clean Sample& 1444 & 23.4 & 1391 & 17.5 & 337 & 1107 & 107.3 & 53  \\
\enddata
\end{deluxetable}
\begin{deluxetable}{lcccccc}
\tabletypesize{\scriptsize}
\tablewidth{0pt}
\rotate
\tablecaption{List of ATCA blazar candidates}
\tablehead{\colhead{3FGL Name}  &   \colhead{Counterpart name}  & \colhead{R.A. radio} & \colhead{Dec. radio} & \colhead{Class count} &     \colhead{Log($\nu^S_\mathrm{peak}$[Hz])} & \colhead{2FGL Name} \\ \colhead{}   & \colhead{} & \colhead{($\arcdeg$)} & \colhead{($\arcdeg$)} & \colhead{} & \colhead{}}
\startdata
%3FGL Name  & Nickname& R.A. & Dec. & Other name  & R.A. radio & Dec. radio & class_counterpart &   redshift & $Syn_{peak}$ &2FGL Name\\
%\hline
J0102.1+0943 & NVSS J010217+094407 & 15.57133 & 9.73622 & BCU II & 14.419 & J0102.2+0943 \\ 
J0437.7-7330 & SUMSS J043836-732921 & 69.65392 & -73.48994 & BCU III & \nodata & J0438.0-7331 \\ 
J0725.7-0550 & NVSS J072547-054832 & 111.44867 & -5.80753 & BCU III & \nodata & J0725.8-0549 \\ 
J0737.8-8245 & SUMSS J073706-824836 & 114.47621 & -82.73703 & BCU III & \nodata & J0737.5-8246 \\ 
J0937.9-1435 & NVSS J093754-143350 & 144.47783 & -14.56414 & BCU II & 17.150 & J0937.9-1434 \\ 
J1016.6-4244 & 1RXS J101620.6-424733 & 154.08650 & -42.78975 & BCU II & 15.600 & J1016.4-4244 \\ 
J1038.0-2425 & NVSS J103824-242355 & 159.59987 & -24.39869 & BCU II & 12.550 & J1038.2-2423 \\ 
J1117.2-5338 & MGPS J111715-533816 & 169.31279 & -53.63783 & BCU II & 14.755 & J1117.2-5341 \\
\enddata
%\tablecomments{}
\label{tab:petrov}
\end{deluxetable}
\begin{deluxetable}{lrrr}
\tablecolumns{4}
\tabletypesize{\footnotesize}
\tablecaption{\label{tab-census}Census of sources}
\tablewidth{0pt}
\tablehead{
\colhead{AGN type}&
\colhead{Entire 3LAC}&
\colhead{3LAC Clean Sample\tablenotemark{a}}&
\colhead{Low-latitude sample}
}
\startdata
{\bf All}&{\bf 1591}&{\bf 1444 }&{\bf182}\\
\\
{\bf FSRQ}&{\bf467}&{\bf 414}&{\bf 24}\\
{\ldots}LSP&412&366&24\\
{\ldots}ISP&47&42&0\\
{\ldots}HSP&3&2&0\\
{\ldots}no SED classification &5&4&0\\

\\
{\bf BL Lac}&{\bf 632}&{\bf 604}&{\bf 30}\\
{\ldots}LSP&162&150&8\\
{\ldots}ISP&178&173&6\\
{\ldots}HSP&272&265&12\\
{\ldots}no SED classification &20&16&4\\
\\
{\bf Blazar of Unknown type}&{\bf 460}&{\bf 402}&{\bf 125}\\
%{\ldots}LSP&198&164&56\\
%{\ldots}ISP&89&79&14\\
%{\ldots}HSP&120&118&33\\
%{\ldots}no SED classification &53&41 & 22\\
{\ldots}BCU I&57&49&11\\
{\ldots}LSP BCU I& 26 & 24 & 8\\
{\ldots}ISP BCU I& 11& 9 & 1\\
{\ldots}HSP BCU I& 13 & 13 & 2\\
{\ldots}BCU I w/o SED classification & 7 & 3 & 0\\
{\ldots}BCU II&346&308&85\\
{\ldots}LSP BCU II & 156 &129 & 39\\
{\ldots}ISP BCU II & 78 & 70& 13\\
{\ldots}HSP BCU II& 107& 105& 31\\
{\ldots}BCU II w/o SED classification &5& 4& 2\\
{\ldots}BCU III&57&45&29\\
{\ldots}LSP BCU III& 16 & 11&9\\
{\ldots}ISP BCU III& 0 &0& 0\\
{\ldots}HSP BCU III &0 &0& 0\\
{\ldots}BCU III w/o SED classification &41 & 34 & 20 \\
\\
{\bf Non-blazar AGN}&{\bf 32}&{\bf 24}&{\bf 3}\\
{\ldots}CSS &2&1&0\\
{\ldots}NLSy1 &5& 5 &0\\
{\ldots}RG &14& 13 &2\\
{\ldots}SSRQ &5&3 &0\\
{\ldots}Other AGN &6&2&1\\
\\
\enddata
\tablenotetext{a}{Sources with single counterparts and without analysis flags. See Section~\ref{sec:classif} for the definitions of this sample.}
\end{deluxetable}
\begin{deluxetable}{llrrcccccccccccc}
\tabletypesize{\tiny}
\tablewidth{0pt}
\rotate
\tablecolumns{15}
\tablecaption{High Latitude ($|b|>10\arcdeg$) 3LAC Full Sample }
\tablehead{
\colhead{3FGL source} &  
\colhead{Counterpart} & 
\colhead{RA} &
\colhead{DEC} &
\colhead{AngSep}&
\colhead{$\theta_{95}$}&
\colhead{Optical} &
\colhead{SED} &
\colhead{log($\nu^{S,meas}_\mathrm{peak}$)} &
\colhead{log($\nu^S_\mathrm{peak}$)} &
\colhead{z} &
\colhead{Prob.} &
\colhead{Rel.} &
\colhead{Rel.} &
\colhead{Compton} 
\\
\colhead{name} &
\colhead{name} &
\colhead{($\arcdeg$)} &
\colhead{($\arcdeg$)} &
\colhead{($\arcdeg$)} &
\colhead{($\arcdeg$)} &
\colhead{ Class} &
\colhead{ Class} &
\colhead{} &
\colhead{} &
\colhead{} &
\colhead{Bay.} &
\colhead{LR$_{RG}$} &
\colhead{LR$_{XG}$} &
\colhead{dominance}
 \\
\colhead{(1)} &
\colhead{(2)} &
\colhead{(3)} &
\colhead{(4)} &
\colhead{(5)} &
\colhead{(6)} &
\colhead{(7)} &
\colhead{(8)} &
\colhead{(9)} &
\colhead{(10)} &
\colhead{(11)} &
\colhead{(12)} &
\colhead{(13)} &
\colhead{(14)} &
\colhead{(15)} 
}
\startdata
J0001.2$-$0748*&PMN J0001$-$0746&0.32510&$-$7.77411&0.042&0.075&BLL&ISP&14.486 & 14.486 &   \nodata &0.9978&0.859&0.00 & \nodata \\
J0001.4+2120*  &             TXS 2358+209 &   0.38502 &  21.22679 &     0.113   &   0.199  &   FSRQ & ISP &     14.163  &   14.486 &   1.10600  &    0.924 &     0.00 &     0.000 & 2.40\\
J0002.2$-$4152*   &   1RXS J000135.5$-$415519  &  0.38642 & $-$41.92367   &   0.137 &     0.174 &  BCU II & HSP &    15.800 &    15.800 &   \nodata  &    0.972  &    0.000 &     0.000 & -0.59 \\
J0003.2$-$5246* &                  RBS 0006  &  0.83121 & $-$52.79103  &    0.017 &     0.065  & BCU II & HSP &    16.850  &   16.850   & \nodata    &  0.998 &     0.000 &     0.900 & -0.52 \\
J0003.8$-$1151         &     PMN J0004-1148 &   1.02048 & $-$11.81622  &    0.076  &    0.114 &  BCU II & LSP &    12.515  &   12.515 &   \nodata  &    0.995 &     0.869   &   0.000 & \nodata \\
J0003.8$-$1151 &               PKS 0001$-$121 &    0.92848 & $-$11.86372 &     0.030 &     0.114 &   BCU I &  \nodata  &    \nodata & \nodata  & 1.30999  &    0.988 &     0.871 &     0.000 & \nodata\\
\enddata
\tablecomments{Column 1 and 2 are the 3FGL and counterpart names, columns 3 -- 4 are the counterpart  J2000 coordinates, column 5  gives the angular separation between the gamma-ray  and counterpart positions, column 6 is the 95\% error radius on the gamma-ray position,  column 7 lists the optical class, column 8 is the  spectral energy distribution (SED) class (depending on the synchrotron peak frequency given in column 9), column 10 is the synchrotron peak frequency corrected for the redshift shown in column 11,  columns 12-14 report the probability for the Bayesian method and the two reliability values, LR$_{RG}$ and LR$_{XG}$ for the radio--gamma-ray match and the X-ray--gamma-ray match respectively.  Column 15 reports $log(CD)$. * refers to sources in the Clean Sample, $^i$ refers to sources whose counterparts have been firmly identified,   $^-$ refers to the lower-probability counterpart in a double association (not propagated to the 3FGL catalog).  The full table is available in the on-line version and at: http://www.asdc.asi.it/fermi3lac/.}\label{tab:clean}
\end{deluxetable}
\clearpage
\begin{deluxetable}{lllllll}
\setlength{\tabcolsep}{0.025in}
\hspace*{40in}
\vspace*{3cm}
%\begin{adjustwidth}{-50em}{}
\tabletypesize{\scriptsize}
\rotate
\tablewidth{0pt}
\tablecaption{Non-blazar objects and misaligned AGN}
\tablehead{
\colhead{Name} &
\colhead{3FGL} &
\colhead{2FGL}&
\colhead{1FGL} &
\colhead{Type} &
\colhead{Photon index} &
\colhead{Notes} }
\startdata
NGC~1218 & J0308.6+0408$^{*}$ & \nodata & J0308.3+0403$^{*}$ & FRI & 2.07$\pm$0.11 & \\
IC~310 & J0316.6+4119$^{*}$ & J0316.6+4119 & \nodata & FRI/BLL & 1.90$\pm$0.14 & \cite{ner10} \\
NGC~1275 & J0319.8+4130$^{*}$ & J0319.8+4130$^{*}$ & J0319.7+4130$^{*}$ & FRI & 2.07$\pm$0.01 & \cite{ngc1275,kat10}  \\
1H~0323+342 & J0325.2+3410$^{*}$ &  J0324.8+3408$^{*}$ & J0325.0+3403$^{*}$ & NLSy1 & 2.44$\pm$0.12 & \\
4C~+39.12 & J0334.2+3915$^{*}$ & \nodata & \nodata & FRI/BLL? & 2.11$\pm$0.17 & \cite{Gio01}  \\ 
TXS 0348+013 & J0351.1+0128$^{*}$ &   \nodata &  \nodata & SSRQ & 2.43$\pm$0.18 &\\
3C~111 & J0418.5+3813 & \nodata & J0419.0+3811 & FRII & 2.79$\pm$0.08 &  \cite{magn,kat11,gra12} \\
Pictor~A & J0519.2$-$4542$^{*}$ & \nodata & \nodata & FRII & 2.49$\pm$0.18 & \cite{bro12,kat11}  \\
\enddata
\tablecomments{
%The five bottom sources were reported in previous catalogs but are missing from 3LAC.  NGC~1275  is 0FGL~J0320.0+4131 and Cen~A = 0FGL~J1325.4$-$4303 \\
%note: () = denotes nearby but not officially associated \\
SSRQ implies FRII. \\
The table includes the 34 non-blazar objects and MAGN at all latitudes associated with 3FGL sources (Cen A Core and Cen A Lobes constitute a single object). $^{*}$ refers to  sources included in the Clean Sample of a given catalog.}
\label{tab:magn}
\end{deluxetable}
\begin{deluxetable}{llrrcccccccccc}
\tablecolumns{14}
\tabletypesize{\tiny}
\rotate
\tablewidth{0pt}
\tablecaption{Low-Latitude ($|b|<10\arcdeg$) Sample}
\tablehead{
\colhead{3FGL source} &  
\colhead{Counterpart} & 
\colhead{RA} &
\colhead{DEC} &
\colhead{AngSep}&
\colhead{$\theta_{95}$}&
\colhead{Optical} &
\colhead{SED} &
\colhead{log($\nu^{S,meas}_\mathrm{peak}$)} &
\colhead{log($\nu^S_\mathrm{peak}$)} &
\colhead{z} &
\colhead{Prob.} &
\colhead{Rel.} &
\colhead{Rel.} \\
\colhead{name} &
\colhead{name} &
\colhead{($\arcdeg$)} &
\colhead{($\arcdeg$)} &
\colhead{($\arcdeg$)} &
\colhead{($\arcdeg$)} &
\colhead{ Class} &
\colhead{ Class} &
\colhead{} &
\colhead{} &
\colhead{} &
\colhead{Bay.} &
\colhead{LR$_{RG}$} &
\colhead{LR$_{XG}$} \\
\colhead{(1)} &
\colhead{(2)} &
\colhead{(3)} &
\colhead{(4)} &
\colhead{(5)} &
\colhead{(6)} &
\colhead{(7)} &
\colhead{(8)} &
\colhead{(9)} &
\colhead{(10)} &
\colhead{(11)} &
\colhead{(12)} &
\colhead{(13)} &
\colhead{(14)} 
}
\startdata
J0012.4+7040 & TXS 0008+704 & 2.88293 & 70.75878 & 0.115 & 0.105 & BCU II & LSP & 13.075 & 13.075 & \nodata & 0.912 & 0.856 & \nodata \\ 
J0014.6+6119 & 4C +60.01 & 3.70330 & 61.29543 & 0.031 & 0.061 & BCU II & LSP & 13.113 & 13.113 & \nodata & 0.997 & 0.976 & \nodata \\ 
J0014.7+5802 & 1RXS J001442.2+580201 & 3.67471 & 58.03404 & 0.009 & 0.055 & BLL & HSP & 16.640 & 16.640 & \nodata & \nodata & \nodata & 0.936 \\ 
J0015.7+5552 & GB6 J0015+5551 & 3.91737 & 55.86226 & 0.018 & 0.043 & BCU II & HSP & 15.791 & 15.791 & \nodata & 0.998 & 0.868 & 0.952 \\ 
J0035.9+5949 & 1ES 0033+595 & 8.96935 & 59.83459 & 0.010 & 0.018 & BLL & HSP & 17.120 & 17.120 & \nodata & 1.000 & 0.935 & 0.979 \\ 
J0047.0+5658 & GB6 J0047+5657 & 11.75179 & 56.96178 & 0.013 & 0.040 & BLL & - & \nodata & \nodata & 0.74700 & 1.000 & 0.910 & \nodata \\ 
J0047.9+5447 & 1RXS J004754.5+544758 & 11.96611 & 54.79579 & 0.010 & 0.060 & BCU II & HSP & 15.896 & 15.896 & \nodata & \nodata & \nodata & 0.890 \\ 
J0102.8+5825 & TXS 0059+581 & 15.69068 & 58.40309 & 0.023 & 0.020 & FSRQ & LSP & 12.725 & 12.941 & 0.64400 & 0.999 & 0.956 & \nodata \\
\enddata
\tablecomments{Column 1 and 2 are the 3FGL and  counterpart names, columns 3 -- 4 are the counterpart  J2000 coordinates, column 5  gives the angular separation between the gamma-ray  and counterpart positions, column 6 is the 95\% error radius on the gamma-ray position,  column 7 lists the optical class, column 8 is the  spectral energy distribution (SED) class (depending on the synchrotron peak frequency given in column 9), column 10 is the synchrotron peak frequency corrected for the redshift shown in column 11,  columns 12-14 report the probability for the Bayesian method and the two reliability values, LR$_{RG}$ and LR$_{XG}$ for the radio--gamma-ray match and the X-ray--gamma-ray match respectively.} \label{tab:lowlat}
\end{deluxetable}
\begin{deluxetable}{llccccccccc}
\setlength{\tabcolsep}{0.025in}
%\HSPace*{-20in}
%\begin{adjustwidth}{-50em}{}
\tabletypesize{\scriptsize}
\tablewidth{450pt}
\tablecaption{Sources from earlier FGL catalogs missing in 3LAC}
\tablehead{
\colhead{Counterpart}  &
\colhead{RA} &
\colhead{DEC} &
\colhead{Optical} &
\colhead{SED} &
\colhead{Redshift} &
\colhead{1FGL}&
\colhead{2FGL}& \\
\colhead{Name} &
\colhead{($\arcdeg$)} &
\colhead{($\arcdeg$)} &
\colhead{Class} &
\colhead{Class} &
\colhead{} &
\colhead{Name} &
\colhead{Name} \\
\colhead{(1)} &
\colhead{(2)} &
\colhead{(3)} &
\colhead{(4)} &
\colhead{(5)} &
\colhead{(6)} &
\colhead{(7)} &
\colhead{(8)} \\
}
\startdata
CRATES J0009+0628 & 2.26638 & 6.47256 & BLL & LSP & \nodata & 1FGL J0008.9+0635&  \\ 
CGRaBS J0011+0057 & 2.87667 & 0.96439 & FSRQ & LSP & 1.492 & 1FGL J0011.1+0050&  \\ 
GB6 J0013+1910 & 3.48509906 & 19.1782456 & BLL & - & 0.473 & & 2FGL J0013.8+1907  \\ 
PKS 0056-572 & 14.6940417 & -56.98675 & BCU & LSP & \nodata & & 2FGL J0059.7-5700  \\ 
PKS 0116-219 & 19.73858 & -21.69167 & FSRQ & LSP & 1.165 & 1FGL J0118.7-2137&  \\ 
TXS 0154-244 & 29.1237238 & -24.2146961 & BCU & - & \nodata & & 2FGL J0156.5-2419  \\ 
S5 0159+723 & 30.8883864 & 72.5483361 & BLL & LSP & \nodata & 1FGL J0203.5+7234 & 2FGL J0203.6+7235  \\ 
B2 0200+30 & 30.93898 & 30.69142 & - & - & \nodata & 1FGL J0203.5+3044&  \\
\enddata
\tablecomments{ Column 1 is the  counterpart name, columns 2 -- 3 are the counterpart   J2000 coordinates, column 4 lists the optical class, column 5 is the  spectral energy distribution (SED) class (depending on the synchrotron peak frequency), column 6 is the counterpart redshift, columns 7--8 show the names in previous {\it Fermi}-LAT catalogues.}\label{tab:missing}
\end{deluxetable}
\begin{deluxetable}{llrcccrrr}
%\setlength{\tabcolsep}{0.025in}
%\HSPace*{-20in}
%\begin{adjustwidth}{-50em}{}
\tabletypesize{\scriptsize}
\tablewidth{0pt}
\tablecaption{3LAC sources: Fluxes (high-latitude sources)}
\tablehead{
\colhead{3FGL source } &
\colhead{Counterpart } &
\colhead{Radio flux} &
\colhead{Radio flag} &
\colhead{X-ray flux$\dagger$} &
\colhead{USNO B1} &
\colhead{SDSS} &
\colhead{$\alpha_{ox}$} &
\colhead{$\alpha_{ro}$}\\
\colhead{name} &
\colhead{name} &
\colhead{(mJy)} &
\colhead{} &
\colhead{($10^{-13}$ erg cm$^{-2}$ s$^{-1}$)} &
\colhead{Vmag} &
\colhead{Vmag} &
\colhead{} &
\colhead{}\\
\colhead{(1)} &
\colhead{(2)} &
\colhead{(3)} &
\colhead{(4)} &
\colhead{(5)} &
\colhead{(6)} &
\colhead{(7)} &
\colhead{(8)} &
\colhead{(9)} \\}
\startdata
J0001.2$-$0748* & PMN J0001$-$0746&  209.17& N&       8.100 &    17.612 &     17.210  &     0.50&       1.38\\
 \nodata&\nodata&\nodata&\nodata&\nodata&\nodata&\nodata&\nodata&\nodata \\
\enddata
\tablecomments{ Column 1 is the 3FGL name, column 2  is the  candidate counterpart name, column 3 is the radio flux measured in the survey indicated in column 4: N for NVSS (1.4 GHz), S for SUMSS (845 MHz),  A for ATCA (20 GHz), P indicates PMN (4.8 GHz) and F indicates FIRST at 1.4 GHz. Column 5 is the X-ray flux between 0.1 and 2.4 keV from the RASS survey  \citep{RASSbright,RASS_FAINT_CAT}, columns 6--7 shows the USNO and SDSS V  magnitudes, respectively.  Columns 8--9 shows the broadband indices between 5000 \r{A}  and 1 keV ( $\alpha_{ox}$) and between 5 GHz and 5000 \r{A} ($\alpha_{ro}$).  
* refers to sources in the Clean Sample. $^-$ refers to the lower-probability counterparts in double associations (not  propagated to the 3FGL catalog). The full table is available in the on-line version and at: http://www.asdc.asi.it/fermi3lac/.}
%\tablecomments{$\dagger$ between 0.1 and 2.4 keV, * refers to sources in the Clean Sample, $^-$ refers to  counterparts not {\bf propagated to} the 3FGL catalog{\bf due to duplicate associations}.  The full table is available in the on-line version and at: http://www.asdc.asi.it/fermi3lac/.}
\label{tab:prob1}
\end{deluxetable}
\begin{deluxetable}{llrcc}
\setlength{\tabcolsep}{0.025in}
%\HSPace*{-20in}
%\begin{adjustwidth}{-50em}{}
\tabletypesize{\scriptsize}
\tablewidth{0pt}
\tablecaption{3LAC sources: Fluxes (low-latitude sources)}
\tablehead{
\colhead{3FGL source } &
\colhead{Counterpart } &
\colhead{Radio flux} &
\colhead{Radio flag} &
\colhead{X-ray flux$\dagger$} 
\\
\colhead{name} &
\colhead{name} &
\colhead{(mJy)} &
\colhead{} &
\colhead{($10^{-13}$ erg cm$^{-2}$ s$^{-1}$)} 
\\
\colhead{(1)} &
\colhead{(2)} &
\colhead{(3)} &
\colhead{(4)} &
\colhead{(5)} 
\\
}
\startdata
J0012.4+7040 & TXS 0008+704 & 639 & N & \nodata \\ 
J0014.6+6119 & 4C +60.01 & 4040 & N  & \nodata \\ 
J0014.7+5802 & 1RXS J001442.2+580201 & 7.7 & N & 104  \\ 
J0015.7+5552 & GB6 J0015+5551 & 85 & N & 152  \\ 
J0035.9+5949 & 1ES 0033+595 & 148 & N & 318  \\ 
J0047.0+5658 & GB6 J0047+5657 & 190 & N  & \nodata \\ 
J0047.9+5447 & 1RXS J004754.5+544758 & 13.9 & N & 31.2  \\ 
J0102.8+5825 & TXS 0059+581 & 849 & N  & \nodata \\
\enddata
\tablecomments{ Column 1 is the 3FGL name, column 2  is the  candidate counterpart name, column 3 is the radio flux measured in the survey indicated in column 4: N for NVSS (1.4 GHz), S for SUMSS (845 MHz),  A for ATCA (20 GHz), P indicates PMN (4.8 GHz) and F indicates FIRST at 1.4 GHz. Column 5 is the X-ray flux between 0.1 and 2.4 keV from the  RASS survey  \citep{RASSbright,RASS_FAINT_CAT}. Parameters in columns 6-9 of Table \ref{tab:prob1}  have been omitted here since they are all blank for this sample. }
%\tablecomments{ $\dagger$ between 0.1 and 2.4 keV, * refers to sources in the Clean Sample, $^-$ refers to counterparts not {\bf propagated to} the 3FGL catalog {\bf due to duplicate associations}. The full table is available in the on-line version and at: http://www.asdc.asi.it/fermi3lac/.}
\label{tab:prob2}
\end{deluxetable}
\begin{deluxetable}{lcccccc}
\setlength{\tabcolsep}{0.025in}
\hspace*{-20in}
\vspace*{-1cm}
%\begin{adjustwidth}{-50em}{}
\tabletypesize{\scriptsize}
\tablewidth{0pt}
\tablecaption{Properties of the 3FGL VHE AGNs. }
\tablehead{
\colhead{3FGL Name} &
\colhead{VHE Name} &
\colhead{Source} &
\colhead{SED} &
\colhead{Redshift} &
\colhead{Spectral} &
\colhead{Variability} \\
\colhead{} &
\colhead{} &
\colhead{Class} &
\colhead{Type} &
\colhead{} &
\colhead{Index} &
\colhead{Index} \\
}
\startdata
J0013.9$-$1853   & SHBL J001355.9$-$185406   &  BL Lac       &  HSP        & 
 0.095                & 1.93$\pm$0.16 & 36.45      \\
J0013.9-1853  & SHBL J001355.9-185406 & BLL & HSP & 0.094 & 1.935$\pm$0.167 & 36.46 \\ 
J0033.6-1921  & KUV 00311-1938 & BLL & HSP & 0.610 & 1.715$\pm$0.035 & 64.62 \\ 
J0035.9+5949$^*$  & 1ES J0033+595 & BLL & HSP & \nodata & 1.898$\pm$0.042 & 69.55 \\ 
J0136.5+3905  & RGB J0136+391 & BLL & HSP & \nodata & 1.696$\pm$0.025 & 62.30 \\ 
J0152.6+0148  & RGB J0152+017 & BLL & HSP & 0.080 & 1.887$\pm$0.103 & 51.76 \\ 
J0222.6+4301  & 3C 66A & BLL & ISP & 0.444 & 1.880$\pm$0.017 & 885.04 \\ 
J0232.8+2016  & 1ES 0229+200 & BLL & HSP & 0.139 & 2.025$\pm$0.150 & 49.16 \\
\enddata
\tablecomments{* refers to low-latitude sources (not in 3LAC).}
\label{tab:GeVTeV}
\end{deluxetable}
\end{document}